\documentclass{article}
\usepackage{graphicx} 
\usepackage{authblk} 
\usepackage{abstract}
\usepackage{tabularx}
\usepackage{amsmath}
\usepackage{url}
\usepackage[hidelinks]{hyperref}
\usepackage{tcolorbox}
\usepackage{caption}
\usepackage{subcaption} 
\usepackage{float}
\usepackage[backend=biber,style=authoryear]{biblatex}
\usepackage[a4paper, total={6in, 8in}]{geometry}

\title{Agent-Based Insight into Eco-Choices: Simulating the Fast Fashion shift}

\author[1,3]{Daria Soboleva}
\author[2,4]{Angel Sánchez}
\affil[1]{Department of Economics, Duke University, Durham, NC 27708, USA}
\affil[2]{Grupo Interdisciplinar de Sistemas Complejos GISC, Departamento de Matemáticas, Universidad Carlos III de Madrid, Leganés, Madrid 28911, Spain}
\affil[3]{Department of Mathematics, Universidad Carlos III de Madrid, Leganés, Madrid 28911, Spain}
\affil[4]{Instituto de Biocomputación y Física de Sistemas Complejos, Universidad de Zaragoza, Zaragoza 50018, Spain}
\affil[ ]{Email: \texttt{daria.soboleva@duke.edu}, \texttt{anxo@math.uc3m.es}}

\begin{document}

\maketitle

\begin{abstract}
    Fashion is a powerful force in the modern world. It is one of the most accessible means of self-expression, thereby playing a significant role in our society. Yet, it is plagued by well-documented issues of waste and human rights abuses. Fast fashion in particular, characterized by its disposable nature, contributes extensively to environmental degradation and CO$_2$ emissions, surpassing the combined outputs of France, Germany, and the UK, but its economic contributions have somewhat shielded it from criticism. In this paper, we examine the demand for fast fashion, with a focus on Spain. We explore the individual decision-making process involved in choosing to buy fast fashion and the role of awareness regarding working conditions, environmental consequences, and education on sustainable fashion in influencing consumer behavior. By employing Agent-Based Modeling, we investigate the factors influencing garment consumption patterns and how shifts in public opinion can be achieved through peer pressure, social media influence, and government interventions. Our study revealed that government interventions are pivotal, with the state's campaigns setting the overall tone for progress, although its success is conditioned by social media and polarization levels of the population. Importantly, the state does not need to adopt an extremely proactive stance or continue the campaigns indefinitely to achieve optimal results, as excessive interventions yield diminishing returns. 

\end{abstract}


\newpage
\section{Introduction}
Consumption of goods and services functions as a medium for expressing and reinforcing our identity to both ourselves and others. Clothing in particular is not just mere utility but a visual language that allows individuals to signal their belonging to certain social groups, subcultures, or communities. This phenomenon reaches its apogee in cultures where consumption became the main way of self-actualization and self-expression. The imperative to continually evolve one's identity through new clothing and the allure of staying fashionable sometimes overshadows concerns about environmental impact, labor conditions, and ethical sourcing. How does this widespread shift in our consumption patterns impact both our world and its inhabitants?

The rise of the fast fashion business model was a response to the growing demand for inexpensive yet trendy clothing. This rapid shift towards lower prices and quality, however, came at the expense of reducing wages, extending the working hours of garment workers, increasing the production of cheaper textiles, and other shortcuts. As a result of these changes in manufacturing practices, the number of times a garment is worn on average has steadily declined, with people accumulating more clothes than ever before.

Fast fashion retailers’ success largely hinges on consumer behavior, fueled by the desire to own more clothing as prices steadily decrease. Although individual actions are often perceived as insignificant compared to the environmental impact of large corporations, studies have highlighted the importance of the consumer's role in European policies (\cite{consumers_matters}). This study aims to investigate the impact of awareness and education about sustainability on individuals' decision-making processes regarding fast fashion purchases. Additionally, we employ Agent-Based Modeling to assess how various influences can alter decision-making and steer public opinion towards more sustainable garment choices. Recognizing that public willingness to endorse and actively participate in initiatives addressing issues like fast fashion is pivotal for progress (\cite{pluralistic_ignorance}), we seek to examine shifts in public opinion through diverse influences such as social media, peer pressure, and governmental intervention.

We model two kinds of agentsets: non-polarized, whose opinions tend to homogenize over time, consistent with classical opinion dynamics models (\cite{consensus_1}, \cite{consensus_2}), and polarized, where agents tend to adhere to their initial beliefs and grow stronger in them. In this case, polarization refers to the processes of driving individual behavior apart. We will come back in detail to the subject of polarization, including a literature review, in Section \ref{polarization_section} below. We find that increased communication generally results in positive changes in concerns, with medium communication levels benefiting non-polarized agents and high communication levels favoring polarized agents. For polarized populations, extreme tolerance levels are less effective, with moderate tolerance yielding the best results. Social media significantly affects non-polarized agents, where moderate pro-sustainability exposure enhances positive changes in concerns, and extreme biases can be detrimental. Conversely, polarized agents show less sensitivity to social media, with lower engagement levels yielding better results. Our analysis of government interventions reveals that while both non-polarized and polarized agentsets respond to pro-sustainability campaigns, the effects are more pronounced in non-polarized agents. Polarized agents show greater resistance to anti-sustainability campaigns and exhibit reduced variability in outcomes. Overall, our simulations suggest that while social media and government interventions shape public concerns and opinions, their effectiveness is significantly influenced by the initial polarization levels and communication dynamics within the population.

The rest of this paper will follow this structure: in Section \ref{literature_overview}, we review existing literature on fast fashion and justify the necessity of this research. Section \ref{data} briefly explores the data we used. In Section \ref{methods}, we provide a detailed description of the model, including the methods employed, an explanation of agent-based modeling, and an overview of the model's structure. Next, Section \ref{results} presents an analysis of models and simulations conducted. Finally, in Section \ref{conclusion}, we summarize the findings, address limitations, and propose avenues for future research.

\section{Literature overview} \label{literature_overview}
The fashion industry ranks as the world's third most polluting sector, surpassing transportation and food retail, and contributing to approximately 8-10$\%$ of our annual carbon footprint (\cite{fashion_pollutes_10_2}). Its environmental impact outweighs that of all international flights and maritime shipping combined (\cite{fashion_pollutes_10}). The Ellen MacArthur Foundation, in collaboration with UNEP, estimates that every second, a truckload of discarded textiles is either dumped in landfills or incinerated. 

Fast fashion denotes the rapid manufacture of inexpensive, substandard clothing often imitating popular styles from fashion labels, renowned brands, and independent designers (\cite{FF_definition}). It has risen to prominence in the fashion industry by selling large volumes of apparel at affordable rates, triggering an unprecedented level of garment consumption. In this section, we will examine fast fashion and its adverse effects on the environment, as well as its impact on the individuals involved in production. We will also explore the cultural dynamics that foster the growth of fast fashion, including consumer behavior.

\subsection{Fashion's environmental impact}\label{env_impact_ff_subsection}

What makes clothing so environmentally harmful? The current fashion business models impact the environment in four primary ways: through the use of cheap materials, outsourcing manufacturing to carbon-intensive locations with extensive transportation needs, high water consumption, and the alarming levels of waste generated. 

\textit{\textbf{Cheap materials}}. To meet the growing demand, garments must be produced rapidly, leading to a reliance on lower-quality materials. Synthetic fabrics like polyester, rayon, and nylon are increasingly favored over natural fibers for their cost-effectiveness. Polyester (PET) is particularly prevalent, making up 51$\%$ of all materials used in the industry. However, it is derived from fossil fuels and may take, according to conservative calculations, up to 200 years to decompose under natural conditions (\cite{PET_decomposition}), unlike natural fibers. Remarkably, the textile industry consumes more PET globally than plastic bottles and other PET products combined (\cite{PET_most_used_in_FF}). Cotton, another commonly utilized material, demands substantial water and pesticide usage during cultivation, and accounts for 26$\%$ of textile production materials (\cite{cotton_water}). Cotton farming often occurs in regions with limited water resources, and relies significantly on the use of chemicals and pesticides. 

\textit{\textbf{Manufacturing locations and Supply Chain}}. The relocation of fast fashion manufacturing to countries with higher carbon footprints, such as India and those in Southeast Asia, is primarily motivated by cost-cutting measures (\cite{supply_chain_main_article}). In these countries, emissions control and regulations are often lax, enabling companies to prioritize profit extraction even at significant environmental expense. However, this shift presents significant challenges to energy supply security and compliance with emissions reduction commitments (\cite{manufacturing_location_1}). Often, garment production involves multiple countries, leading to increased logistical steps between processes driven by economic considerations (\cite{diff_locations_manufacturing}). Additionally, there is a growing reliance on air cargo for transportation, significantly amplifying environmental impact. 

\textit{\textbf{Water consumption}}. The fashion industry stands as the leading contributor to freshwater pollution globally (\cite{fashion_pollutes_10}), accounting for 20$\%$ of the world's wastewater (\cite{supply_chain_main_article}). With over 1900 chemicals identified in textile production processes (\cite{chemicals_in_water}), industrial wastewater from this sector often contains hazardous dyes and other pollutants detrimental to aquatic life and human health. Additionally, the physicochemical properties of this wastewater can prevent its biodegradation (\cite{cannot_degrade}). 

\textit{\textbf{Textile waste}}. Many assume that once a garment is purchased, its impact on the planet ceases. However, this could not be further away from the truth. Currently, vast clothing piles occupy around 5$\%$ of landfill space, predominantly in lower to middle income countries (LMICs) (\cite{space_in_lmics}). Unsold and 90$\%$ of the donated items eventually contribute to solid waste (\cite{landfill}), cluttering waterways, green spaces, and public parks, thereby posing additional environmental and health risks in LMICs.

\subsection{Fast fashion and working conditions}\label{ff_work_conditions_subsection}

Beyond its detrimental environmental impacts and the adverse effects on the quality of life for populations in LMICs, fast fashion has attracted global attention for its exploitative practices and the poor conditions and wages of garment workers. While working conditions have improved in the developed world, the labor rights abuses and violations have not disappeared; instead, they have shifted overseas, where lack of proper governance and regulation, and corruption prevent the implementation of appropriate measures.

On a positive note, the fast fashion sector has contributed to overseas income, foreign exchange earnings, women's empowerment, and the overall export value and GDP in countries such as Bangladesh (\cite{increase_GDP_Bangladesh}) and Ethiopia (\cite{increase_GDP_Ethiopia}). However, the notion of economic development transcends merely raising the per capita income of industry workers. Coupled with increasing income inequality, the working class often remains trapped in poverty, facing job insecurity, meager wages, extended work hours, limited access to public services, substandard health care, and dismal living and working conditions (\cite{uzb_papers}). They also endure a lack of legal rights and face physical and mental threats to their economic and social well-being.

Although consumers are increasingly concerned about unethical practices, this sentiment often fails to translate into action. The avoidance of personal sacrifice frequently leads to a significant discrepancy between attitudes and actions.

\subsection{Throw away culture} \label{throw_away_culture_subsection}

The evolution of the fashion business model has been closely related to shifts in consumer attitudes toward clothing. The frequent updates of styles and affordability have fostered a culture of impulse buying and a perception of clothing as easily disposable. Fast fashion brands now introduce up to 52 seasons per year (\cite{52seasons}), compared to the traditional four seasons, capitalizing on limited collections that quickly go out of style and the planned obsolescence of their products due to their poor quality. 

Additionally, several barriers hinder the adoption of more sustainable shopping habits. Sustainable clothing is often perceived as more expensive than fast fashion items, and there persists a belief that sustainably produced clothing lacks attractiveness and is thus considered unfashionable. Social norms also play a significant role in shaping our shopping behavior, with normative expectations influencing how we perceive our peers' expectations of us. Consequently, even when individuals express concerns about the environmental and social impacts of fast fashion, these concerns do not necessarily imply changes in their shopping habits. This phenomenon, known as the ``Ethical Purchasing Gap" (\cite{attitude_behavior_gap}), suggests that in fashion purchases, ethical considerations may have some influence, but factors like color and style are likely to be more significant determinants.

\section{Data} \label{data}

For this study, we utilize data gathered by Silvia Blas Riesgo for the study \textit{Drivers and barriers for sustainable fashion consumption in Spain: a comparison between sustainable and non-sustainable consumers} (\cite{Drivers_and_barriers}). This dataset comprises over 80 questions pertaining to 1067 Spanish respondents’ clothing preferences, shopping habits, environmental concerns, awareness of working conditions, demographic information, and more. Spanish society presents a particularly interesting case for two reasons: clothing is a paramount form of self-expression and an important aspect of social life in many parts of the country, and Spain is home to Inditex — a major industry player contributing significantly to the country's economic growth. The survey responders identified predominantly as women (80$\%$), and half of the responders were of the age category 18-24. This turns out to be beneficial for our study, since these are the two categories of people most likely to engage in overconsumption of clothing (\cite{consumers_age}).

\section{Methods}  \label{methods}

\subsection{Purpose}

We aim to study garments consumption patterns, identify barriers to adopting more sustainable shopping practices, assess responses to educational interventions, and evaluate the impact of policies and external influences. To do so, we employ Agent-Based Modeling (ABM), which is a computational modeling technique utilized to simulate and analyze the behavior and interactions of diverse individual agents at a micro-level, and the resultant outcomes at the macro-level of a system (\cite{ABM_description}). 

The model is implemented in the multi-agent programmable environment NetLogo 6.3.0. A copy of the model with supplementary information can be downloaded from the model library of the CoMSES Net website (\cite{soboleva2024agent}). 

Two important aspects of ABM are choosing a decision-making model for individual agents and initializing the agents with real-world data.

\subsection{Decision-making model}
The initial step involves identifying the microdrivers of behavior. These represent the fundamental factors influencing individuals’ shopping habits, and are determined by using a linear regression analysis. Each agent within our model will be categorized based on their likelihood of purchasing from a fast fashion brand. The objective of this phase is to uncover the traits, preferences, and characteristics that forecast a higher or lower probability of engaging in fast fashion consumption.

From the dataset, we identify several predictor variables and one response variable, which are described in detail in Appendix \ref{decision_making_appendix}. 

The predictor variables of choice are: 
\begin{itemize}
    \item Environmental concerns;
    \item Working conditions awareness;
    \item Shopping frequency;
    \item Education on the topic of sustainable fashion;
    \item Normative expectations;
    \item Trust in sustainable companies;
    \item Access to sustainable fashion.
\end{itemize}

The response variable of choice is the overall probability to purchase fast fashion. It is important to note that the response variable is not the inverse of the probability of purchasing sustainable clothing, but rather agent's view on purchasing fast fashion. We acknowledge that transitioning from high rates of purchasing fast fashion clothing to high rates of purchasing sustainable clothing is not desirable, since the overconsumption of any sort of clothing is in itself unsustainable. Instead, we aim to model individual's likelihood in buying fast fashion clothing as opposed to not buying, buying from a sustainable brand, shopping at a second-hand store, or any other alternative. After conducting the linear regression analysis, we were able to find that 37$\%$ of variability in the response variable were caused by the changes in the predictor variables ($R^2=0.37$). The resulting linear regression formula that will be used for agent's decision making process is given by Eq. (\ref{probability_buy_ff}).

\begin{equation}
    A_p = b_0 + b_1 \cdot A_{sex} + b_2 \cdot A_{age} + b_3 \cdot A_{env} + b_4 \cdot A_{exp}  + b_5 \cdot A_{wca} + b_6 \cdot A_{know} + b_7 \cdot A_{trust} + b_8 \cdot A_{access} + b_9 \cdot A_{freq},
    \label{probability_buy_ff}
\end{equation}

where $A_i \in [0,1]$ are agents' attributes and their corresponding definitions can be found in Table \ref{tab:variables_lin_regression}.  The coefficients come from the linear regression analysis and can be found in Table \ref{tab:results_lin_regression}. Variables not considered in this analysis, detailed in Appendix \ref{decision_making_appendix}, were discarded due to a lack of significant correlation between changes in their values and the response variable. These include: level of education, perceived consumer effectiveness, income. Moreover, while we do include normative expectations and attitude, we do not include behavioral control as suggested by the Theory of Planned Behavior (\cite{TPB}). This choice is due to two reasons. Firstly, one of most common limitation brought up against buying sustainable clothing is its high cost. However, as further explained in Appendix \ref{decision_making_appendix} and shown in Table (\ref{tab:results_lin_regression}), neither income nor lack of access to sustainable cloths were significantly correlated with being a sustainable shopper. Secondly, the behavior we aim to affect is not purchasing sustainable clothing, but rather \textit{not purchasing} clothing from fast fashion brands. This can be seen as buying sustainable clothing, shopping second hand, or simple reducing how much an individual shops for clothing. This is consistent with our definition of the outcome variable. 

\begin{table}
        \centering
	\begin{tabular}{|c|c|}
		\hline
		\textbf{Coefficient name} & \textbf{Value} \\
            \hline
		$A_p$ & Probability to buy fast fashion \\
            $A_{sex}$ & Sex  \\
            $A_{age}$ & Age \\
            $A_{env}$ & Environmental concerns \\
            $A_{exp}$ & Normative expectations \\
            $A_{wca}$ &  Working conditions awareness \\
            $A_{know} $& Education on sustainable fashion \\
            $A_{trust}$ & Trust in companies \\
            $A_{access}$ & Access to sustainable brands \\
            $A_{freq}$ & Shopping frequency \\
            \hline
	\end{tabular}
        \caption{Variables for Eq. (\ref{probability_buy_ff}).}
        \label{tab:variables_lin_regression}
\end{table}

\begin{table}
        \centering
	\begin{tabular}{|c|c|c|}
		\hline
		\textbf{Coefficient name} & \textbf{Value} & \textbf{Predictor variable} \\
            \hline
		$b_0$ & 0.7450 & constant \\
            $b_1$ & -0.0101 & Sex  \\
            $b_2$ & 0.0200 & Age \\
            $b_3$ &  -0.0179 & Environmental concerns \\
            $b_4$ & -0.0488 & Normative expectations \\
            $b_5$ & -0.1783 & Working conditions awareness \\
            $b_6 $& -0.1414 & Education sustainable fashion \\
            $b_7$ &  0.0320& Trust in companies \\
            $b_8$ & 0.0360 & Access to sustainable brands \\
            $b_9$ & 0.2181 &Shopping frequency \\
            \hline
	\end{tabular}
        \caption{Coefficients of the linear regression}
        \label{tab:results_lin_regression}
\end{table}

\subsection{Initializing the agentset} \label{initializing_agentset}

Agents are initialized based on the information from the dataset used for the linear regression. It is used to create 1050 agents. Once the agentset has been initialized, we create cliques (close friends) and distant links (acquaintances) according to the mechanism described below in Section \ref{initializing_non_polarized}.  The attributes each agent is assigned are seen in Table \ref{tab:variables_lin_regression} and are those used in the decision-making model. Aside from these, each agent is randomly assigned three additional static variables that represent their levels of susceptibility. Each of these are within the range of $[0.1,0.9]$, where 0.1 indicates low susceptibility and 0.9 - high susceptibility

\begin{enumerate}
    \item \textbf{Peer influence susceptibility}, $S_{pp}$. This variable determines how easily an agent's opinion is influenced by other agents in its neighborhood. A higher susceptibility value indicates that the agent is less influential on others. Conversely, if an agent has significant influence over its peers, its susceptibility to their influence will be lower. This represent the phenomenon seen in real-life interactions, where lower susceptibility tends to correlate with higher influentiality, and vice versa (\cite{suscept_infl}).

    \item \textbf{Social media susceptibility}, $S_{sm}$. This value denotes the degree to which an agent is susceptible to social media and its influence. In this simulation, we assume that all agents utilize some form of social media on a daily basis. The level of susceptibility can also be seen as proportional to the frequency of usage of social media.

    \item \textbf{Government influence susceptibility}, $S_{gov}$. This variable determines the extent to which an agent is influenced by government initiatives and interventions, such as campaigns and education efforts. Different levels of susceptibility are primarily associated with political affiliations: agents with higher susceptibility represent individuals whose political identity aligns with the governing party, to varying degrees. The opposite is true for agents with low government susceptibility, although these can also represent agents with apolitical stances.

\end{enumerate}

\subsection{Agent-Based Model}

We now move onto describing the ABM. There are three key components to this model: peer interaction, social media influence, and government interventions. All three can influence three agents' attributes: environmental concerns, working conditions awareness, and education on the topic of sustainable fashion. Additionally, the government can influence agents' trust in companies claims regarding sustainability. In this section, we describe each for of influencing present in the model.

\subsubsection{Peer influence. Non-polarized agentset} \label{initializing_non_polarized}

The first kind of agentset is titled ``non-polarized" and corresponds to a set of individuals whose opinion on topics tends to homogenize over time because these topics are not perceived as controversial or polarizing ones (\cite{opinions_average_out}). In both non-polarized and polarized agentset, agents are programmed to interact with a subset of the agents in their neighborhood during each time step. Every agent is set to have 5 close friends, which forms their inner circle, and 10 acquaintances, which forms their outer circle. Although in real-life individuals tend to have more connections (\cite{Friend_group_size_1}, \cite{Friend_group_size_2}), we base our choice of number of connections on the limited size of unique agents (1050) to avoid overconnectedness. The inner circle, comprised of agent's close friends, is a fully interconnected clique, while the acquaintances are of an agent are not necessarily connected to each other.

Our arrangement can be viewed as a random network with embedded cliques, that builds dense local groups and adds random weak ties to other agents. The structure of our model shares the small-world properties of the Watts–Strogatz network (\cite{Watts-Strogartz}) — high clustering and short average path lengths. We explicitly embed agents in fully connected cliques and link them to randomly selected others, which allows us to capture the realistic duality of social ties: dense, cohesive friend groups alongside sparse, long-range connections. Unlike Watts–Strogatz, which imposes uniform local connections, or Kleinberg’s model based on spatial proximity (\cite{Kleinberg}), our network explicitly reflects community structure, making it especially suitable for modeling social behavior within clustered populations. Fashion choices often reflect a clustered population dynamic, where individuals are significantly influenced by their immediate social groups or ``cliques." The patterns introduced by out network structure reflect how individuals frequently look to their immediate social circles for cues on fashion choices, leading to the formation of tightly knit groups with similar fashion preferences.

Hence, each agent has a total of 15 connections. Agents interact with their network at every time step. To introduce variability, we allow agents to interact with $10\pm a$ friends, where $a\sim  U[1,4]$. In this model, we assume that an agent is equally likely to interact with their close friends as with their acquaintances. Additionally, we assume that each interaction, whether it is with a close friends or an acquaintance, has an equal effect on the agent's perception of their opinion. While people may value close friends' opinions in real life on some topics, it's unclear if this holds in fashion dynamics. Research suggests individuals tend to follow more influential figures, not necessarily close friends (\cite{law_of_the_few}, due to the expert effect (\cite{social_influence}). We account for influential agents with the attribute ``Peer influence susceptibility," depicted in Section \ref{initializing_agentset}. Moreover, we avoid giving more weight to close friends' opinions to prevent excessive clustering, which is already introduced by the presence of cliques. Similarly, we do not give more weight to interactions with agents whose opinions align in the non-polarized setting. This differs from previous research that combined agent-based modeling and environmental concerns (\cite{abm_environment}), where influence of individuals on each other is a function of their similarity in environmental identity. However, we do introduce an analogous mechanism in polarized agentsets, described in Section \ref{polarization_section}.

Once each agent's neighborhood is determined, the agent's updated opinion after each interaction is given by Eq. (\ref{peer-non-pol-eq}).

\begin{equation}
    A_{t,i} = (1-S_{pp}(A_i))\cdot A_{t-1,i} + \frac{S_{pp}(A_i)}{|C|}\cdot \sum\limits_{j=1}^{|C|}  \left[ \frac{1}{3}A_{t-1,j} + \frac{2}{3}B_{t-1,j}\right],
    \label{peer-non-pol-eq}
\end{equation}
where $A_{t,i}$ stands for agent's $i^{\text{th}}$ updated opinion on a topic at time $t$, $S_{pp}(A_i)$ represents agent's susceptibility towards others, $ A_{t-1,i}$ represents their opinion at the previous time step, $C = \{A_j: A_j\in peers(A_i)\}$ is the set of peers the agent interacted with at previous time step $t$, and $A_{t-1,j}$ and $B_{t-1,j}$ represent peers' opinion and behavior at a previous time step, respectively. 
It is important to note that an agent can only be influenced by those in its neighborhood whose susceptibility is lower than that of the agent. In other words, an agent is only influenced by more influential agents that the agent themselves is (\cite{susceptibilit-influential}). On the one had, this allows us to introduce the concept of assimilation: individuals tend to adopt the culture of the dominant agents. The concept of assimilation is further introduced with the state's influence variable, which is described below. On the other hand, this allows us to incorporate an important social phenomenon: inexorable individuals, which are not easily persuaded, moved, or affected by normative expectations. 
Additionally, both peers' behavior and attitudes are included in this equation ($A_{t-1,j}$ and $B_{t-1,j}$, respectively). This decision was based on the observation that individuals tend to be influenced twice as much by conformity with peers' behavior as by normative expectations (\cite{attitude_vs_behavior}). In other words, what one's peers do matters more than one's expected behavior. This choice of opinion updating mechanism is based on the foundational model of opinion dynamics introduced by Morris H. DeGroot in his 1974 paper \textit{Reaching a consensus} (\cite{consensus_2}), where a group of individuals repeatedly update their beliefs by averaging the beliefs of others in their social network. In particular, it closely resembles another well-knows framework in opinion dynamics, the Friedkin–Johnsen (FJ) model, which is an extension of DeGroot's model. The FJ model allows agents to retain some of their initial opinions.

\subsubsection{Peer influence. Polarized agentset} \label{polarization_section}

The second kind of agentset we explored is titled ``polarized." Our goal is to explore peer influence in a scenario where the population is polarized, which mean that it does not homogenize their views over time, but instead clusters around multiple opinions. Often times, consensus cannot be achieved in real-world scenarios regarding polarized topics for a variety of reasons. For instance, individuals' initial internal opinions, unlike their expressed opinions, can remain unchanged by social interaction (\cite{opinion_do_not_homog}). Moreover, polarization seems to burst especially in public discussions evolving around politically and ethically controversial issues (\cite{pol_polarization_1}, \cite{pol_polarization_2}). In certain societies and countries, fast fashion, its relation to the environment and unethical practices already constitutes a polarized topic. Therefore, although classical opinion dynamics models suggest that increased interaction rates would eventually lead to a consensus (\cite{consensus_1}, \cite{consensus_2}), even on controversial issues, this has been challenged by vast empirical evidence of opinion polarization. We account for such developments by introducing a polarized agentset. This mechanism similar to existing ABM models on environment adjacent topics (\cite{abm_environment}), where agents are influenced by others based on similarities in identity. 

The equation used for social interactions in a polarized society will be similar to that for a non-polarized society. However, the difference lies in the introduction of a tolerance threshold (denoted by $\tau \in [0.05, 0.50]$), which represents how tolerant agents are toward opinions that deviate from theirs by more or less than the threshold. This is a further extension of the Friedkin–Johnsen framework. The tolerance threshold we introduce results in a opinion dynamic similar to that of the Bounded Confidence Model (\cite{polarization_threshold}), where individuals are influenced only by others whose opinions differ by less than a specified threshold. We provide more details on the modified opinion-updating mechanism in the Appendix \ref{agent_based_details_appendix}. In summary, a lower threshold indicates less tolerance towards differing opinions. When an agent encounters an opinion that exceeds their tolerance threshold, the interaction becomes polarizing rather than unifying.

\subsubsection{Social media influence.}

Social media is treated as a single external entity. While various social media platforms exist, we simplify this complexity by considering social media as a unified external influence. It is important to clarify that from now on, when exposure to or engagement with social media is mentioned, it pertains solely to exposure on the topics of interest (environment, working conditions, sustainability, etc.) and does not refer to all social media usage. 

In real life, social media platforms adapt to each user, tailoring content based on their preexisting preferences and inclinations. Content one sees is content they already consume or are likely to consume. In our model, we implement such mechanisms by incorporating a feedback loop between the agent's opinion and their social media platform. It is recognized that social media tends to increase polarization by creating ``echo chambers" that limit exposure to information contradicting preexisting beliefs (\cite{pol_sm_2}), and this is particularly the case for politicized topics (\cite{pol_sm_3}). Moreover, social media can be biased towards pro-fast fashion or anti-fast fashion, which is something we consider in this model. Our choice to add a bias term rests on the fact that social media is found to both promote consumption (\cite{sm_consumption1}, \cite{sm_consumption2}), but can also be a powerful tool that can be used to promote sustainable habits (\cite{sm_sust1}, \cite{sm_sust2}, \cite{sm_sust3}). As a result, we derive a function that takes an agent's current opinion as an input an returns an ``influence" opinion to the agent. This feedback loop is given by Eq. (\ref{feedback_sm_equation}).

\begin{equation}
    SM\left(A_{t-1}\right) = b \cdot\left(A_{t-1}\right)^3 
    + \frac{-3 \cdot b}{2}\left(A_{t-1}\right)^2 + \frac{3 \cdot b}{4}\left(A_{t-1}\right)
    + \left(\frac{1}{2}-\frac{b}{8}\right) + \beta,
    \label{feedback_sm_equation}
\end{equation}

were $b=50\cdot S_{sm}$, $S_{sm}$ stands for agent's susceptibility towards social media, and $\beta \in [-0.30, 0.30]$ is the bias. The rationale behind the choice of function for this feedback loop is detailed in Appendix \ref{agent_based_details_appendix}. The output varies for each agent's opinion, reflecting the individual calibration of social media platforms. Once we have obtained the social media influence level from the function above, we can find the agent's opinion after using social media, which is given by:

\begin{equation}
    A_{t} = (1-S_{sm})\cdot g(A_{t-1}, SM)\cdot A_{t-1}  + S_{sm}\cdot f(A_{t-1}, SM)\cdot SM,
    \label{social_media_influence}
\end{equation}

where $A_{t-1}$ represents agent's opinion at the previous time step, $A_{t}$ is the updated opinion, $S_{sm}$ stands for social media susceptibility of the agent, and $SM$ is the opinion that social media obtained from the feedback loop. The rationale behind this choice of function is detailed in the Appendix \ref{agent_based_details_appendix}.

\subsubsection{Government interventions.}

The government, or state, is considered a single external entity in this model, with one primary goal: to influence the population's opinions and concerns regarding topics of interest (environment, working conditions, sustainability, trust in companies.) Unlike social media, the government does not tailor campaigns to individual agents based on their current views. Instead, it shares the same information with the entire population. 

In this model, the government is designed as a ``smart" entity aiming for re-election. This means that it promotes opinions similar to the average opinion of the population, which it uses as an input to the government feedback loop, shown in Eq. (\ref{gov_feedbacl_eq}).

\begin{equation}\label{gov_feedbacl_eq}
    GOV(A(tot)_{t-1})=\zeta \cdot A(tot)_{t-1},
\end{equation} 

where $A(tot)_{t-1}$ stands for the average view of the entire population, and $\zeta$ is a parameter set by the user that falls within the range of $[0.5, 1.5]$, where $\zeta = 0.5$ signifies a strong anti-sustainability stance, $\zeta=1$ represents neutrality, and $\zeta=1.5$ indicates strong pro-sustainability views. After the state's ``opinion" is determined by Eq. (\ref{gov_feedbacl_eq}), it is disseminated to agents. The formula representing the change in an agent's opinion after ``interacting" with governmental influence is given by Eq. (\ref{gov_influence}).
\begin{equation}
    A_{t} = (1-S_{gov}) \cdot g(A_{t-1}, GOV) \cdot A_{t-1} + S_{gov} \cdot f(A_{t-1}, GOV) \cdot GOV,
    \label{gov_influence}
\end{equation}

where $GOV$ is the promoted opinion calculated with Eq. (\ref{gov_feedbacl_eq}), $S_{gov}$ is agent's susceptibility to government's interventions, and $A_{t-1}$ is the opinion of an agent at time $t-1$. The functions $g$ and $f$ are exactly the same as described above for social media, and Eq. (\ref{gov_influence}) is identical to Eq. (\ref{social_media_influence}) explained in Appendix \ref{agent_based_details_appendix}. Similar to social media, we aimed to prevent drastic changes in agents' opinions and thus chose a more sophisticated function that allowed for a slower evolution of their viewpoints. Additionally, we implement a function that aims to mimic the diminishing impact of repeated exposure, a phenomenon known as ``campaign fatigue" (\cite{campaign_fatigue}). This will affect agents' susceptibility to government, $S_{gov}$, in a manner represented by Eq. (\ref{decay}).

\begin{equation}
    S_{gov, t} = S_{gov, t-1} \cdot \exp{(-0.00125\cdot T)},
    \label{decay}
\end{equation}

where $S_{gov, t-1}$ is agent's susceptibility at previous time step, $S_{gov, t}$ is that same agent's susceptibility at current time step, and $T$ is the number of weeks (which consist of 7 time steps) that have past since the beginning of the simulation.

\subsection{Structure overview and scheduling}

Once agents and their friend groups are initialized, the simulation begins. It is set to run for 500 time steps, which was chosen as the limit because further iterations provided no qualitative differences in outcomes. Each time step, agents deterministically interact with some of their neighbors and social media once, and stochastically interact with government campaigns. It is important to note that these discrete time steps that do not necessarily correspond to real-world units such as days. This limitation reflects a common issue in ABMs known as \textit{temporal granularity}, and the lack of longitudinal data prevents us from calibrating the model’s time scale. Consequently, the timing should be interpreted qualitatively rather than quantitatively. At each time step, we measure the agent's probability to purchase fast fashion as the variable of interest, which does not necessarily correspond to the act of purchasing, but should rather be viewed as an individual's acceptance of fast fashion and likelihood of purchasing it when faced with the decision. More information on the simulation can be found in Appendix \ref{simulations_appendix}.

\section{Model settings} \label{models}

We considered several different scenarios, and the resulting models can be grouped into three sets. In the first set, we focus on communication and consequent peer pressure. In the second set, we study the different effects of social media influence. In the third set, we look at government interventions. 

Peer pressure was studied both in the context of non polarized and polarized agentsets. We explore how different levels of communication (determined  by the sharing threshold $\delta \in [0.05, 0.5]$) impacts average attributes of interest.  For polarized agentsets, we also study how different levels of tolerance ($\tau$) towards members of the group affects the peer pressure mechanism. Models in this set are denoted by letter A.

Social media has two parameters to work with: the portion of agents that get exposed to social media (determined by $\sigma \in [0.05, 0.5]$), which is based on their levels of concerns) and social media bias ($\beta$). We study both, in particular in combination with increased communication. Models in this set are denoted by letter B.

Lastly, we look at the effects of government interventions taking into account the possible range of different stances (determined by the value of $\zeta \in [0.5, 1.5]$). We also study the resilience agents show after campaigns are halted, and the effect that social media bias has on effectiveness of campaigns. Models in this set are denoted by letter C.

Every model is characterized by a set of parameters. The parameters and their definitions can be found in Table \ref{tab:parameters_used_in_models}. Please refer to the Appendix for a more detailed explanation of these parameters. 

\begin{table}[h!]
    \centering
    \setlength{\tabcolsep}{3pt} 
    \begin{tabularx}{\linewidth}{|c|X|X|}
        \hline
        \textbf{Parameter}  &  \textbf{Description}  & \textbf{Range of values} \\
        \hline
        $\delta$ & Communication threshold & 0.05 (low communication among agents) - 0.5 (high communication) \\
        $\tau$ & Tolerance threshold & 0.05 (low tolerance towards others) - 0.5 (high tolerance, no polarization) \\
        $\sigma$ & Exposure to social media on topics of interest & 0.05 ($10\%$ of agents are exposed) - 0.5 ($100\%$ of agents are exposed) \\
        $\beta$ & Social media bias & (-0.30) (pro-fast fashion bias) - (+0.30) (pro-sustainability bias) \\
        $\zeta$ & Government state & 0.5 (anti-sustainability) - 1.5 (pro-sustainability)\\
        \hline
    \end{tabularx}
    \caption{Parameters used in the models}
    \label{tab:parameters_used_in_models}
\end{table}

\section{Results} \label{results}

In this section, we outline the settings of the models used in this project and present the results. These results are consistent across multiple runs, and although the reported numbers correspond to a single run, they are representative of the average outcomes. Additionally, the code and sample data — reflecting the original data’s distribution — are available on the CoMSES Net website. The section is divided into three parts, each focusing on one mechanism of influence and its interaction with other influences.

\subsection{Set A: peer pressure}

(A1) \fbox{$\delta$: 0.1, 0.3, 0.5,  $\tau = \text{N/A}$, $\sigma = 0.1$, $\beta = 0$, $\zeta = \text{N/A}$} \\

We begin by looking at the communication levels and the effect they have on concerns. For the non polarized agentset, we run a baseline model  with different levels of communication. We find that increased communication positively impacts the net changes in individuals' average concerns, supporting previous research findings (\cite{communication2}, \cite{communication3}, \cite{communication4}). Peer pressure is essential for behavioral change, as behaviors are not solely based on individual preferences but are reinforced by social expectations. Public willingness to accept, support, and actively participate in social, cultural, economic, and political changes is crucial for effective mitigation efforts. The graph showing changes in the probability of purchasing fast fashion can be found in Fig (\ref{fig:graph0}), while more information on changes in concerns can be found in Fig. (\ref{fig:graphA1}) in the Appendix. Our findings for the non-polarized agentset are consistent with DeGroot's model, where the average opinion converges to a consensus. However, it differs from the FJ model results, which postulates that an equilibrium is never reached. In our case, there are not enough stubborn agents and variability in initial opinions to reach results typical for an FJ model. \\

(A2) \fbox{$\delta = 0.1, 0.3, 0.5$, $\tau = 0.15$, $\sigma = 0.1$, $\beta = 0$, $\zeta = \text{N/A}$}\\

We conducted the same experiment with a polarized agent set. Similar to the non-polarized case, we observed an overall improvement in population concerns with increased communication levels. Interestingly, a medium level of communication ($\delta=0.3$) had a more positive effect on the non-polarized agent set than on the polarized one, whereas the highest communication level ($\delta=0.5$) had the opposite effect. The graph of changes in probabilities to purchase fast fashion is depicted in Fig. (\ref{fig:graph0}), while changes in concerns can be found in Fig. (\ref{fig:graphA2}) in the Appendix. 

Notably, in both simulations, the changes in environmental concerns are negative across all communication levels. This may be due to the initially high average environmental concerns at the start of the simulation. However, this high level of concern does not translate into corresponding actions. Since concerns are influenced by both opinions and behaviors, the general lack of action reduces overall concerns because peers' behaviors impact normative expectations as much as, if not more than, their opinions. Moreover, the variability in opinions increases with the increase of the value of $\delta$, which expected given the polarizing nature of the agentset. (Fig. (\ref{fig:graphA2_vars}), Appendix.) 

\begin{figure}[ht]
\centering
\includegraphics[width=\textwidth]{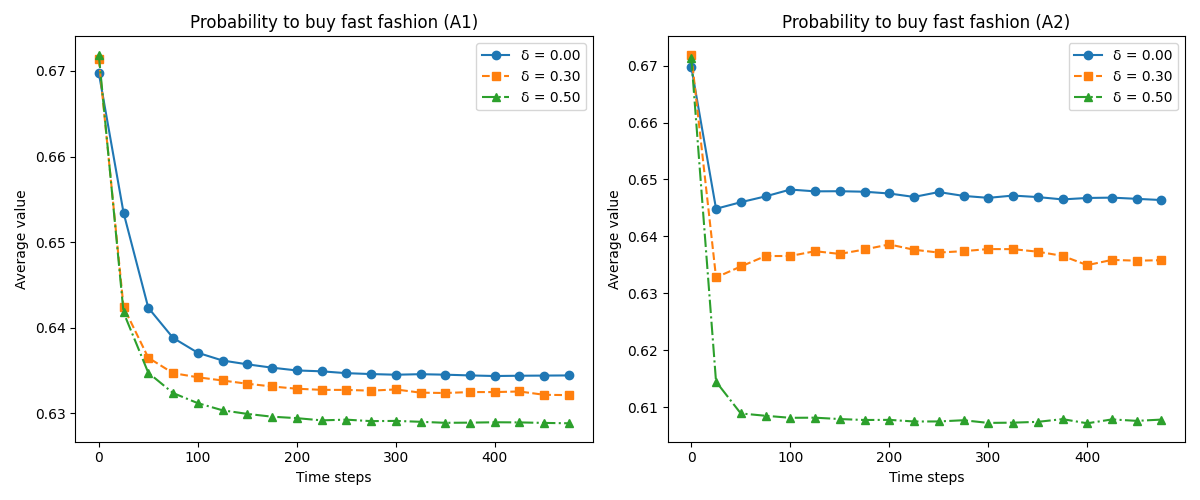}
\caption{Changes in average values of probabilities for Models (A1) and (A2)
\textit{Source: our simulations.}}
\label{fig:graph0}
\end{figure}

(A3) \fbox{$\delta = 0.1, 0.3, 0.5$, $\tau = 0.15$, $\sigma = 0.1$, $\beta = 0$, $\zeta = \text{N/A}$}\\

Additionally, we studied the different levels of tolerance within the polarized agent set and their impact on communication. As the tolerance value ($\tau$) approaches 0.5, the agent set becomes less polarized, effectively becoming non-polarized when $\tau$ reaches its maximum value.  We found that very high or very low tolerance levels are not constructive for communication (Fig. (\ref{fig:graphA3}), Appendix.) Specifically, a tolerance level of $\tau=0.15$ yielded the best results in terms of desired changes. This suggests that extremely low tolerance ($\tau=0.05$) is detrimental to change, as it makes agents more reluctant to alter their opinions. These results reinforce previous findings on how small changes to how individuals behave or how much they value similarity can shift collective dynamics.

\begin{figure}[ht]
\centering
\includegraphics[width=\textwidth]{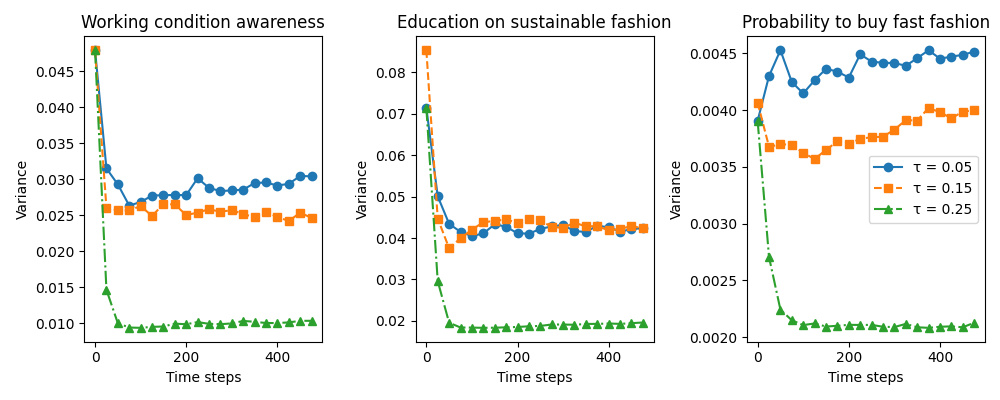}
\caption{Variances of average values for Model (A3)
\textit{Source: our simulations.}}
\label{fig:graphA3_vars}
\end{figure}

Furthermore, very low tolerance results in high opinion variability, as seen in Fig. (\ref{fig:graphA3_vars}). This is expected because agents with low tolerance adopt more extreme opinions and become trapped in echo chambers. High opinion variability indicates that agents are more dispersed in their attitudes, leading to clustering within the polarized agent set, a phenomenon not observed in the non-polarized agent set. Conversely, higher tolerance levels ($\tau=0.25$ and above) lead to significant homogenization of opinions. 

We can conclude that increased communication positively impacts average concerns for both non-polarized and polarized agentsets, with medium communication levels ($\delta=0.3$) being more beneficial for non-polarized agents and high communication levels ($\delta=0.5$) for polarized agents. For the polarized agentset, extreme tolerance levels are not constructive for communication, with $\tau=0.15$ yielding the best results. Very low tolerance ($\tau=0.05$) leads to high opinion variability and clustering, while higher tolerance levels ($\tau=0.25$ and above) result in significant homogenization of opinions.

\subsection{Set B: social media influence}

Social media plays a large role in shaping people's opinions nowadays. In this study, we focus on how it can promote or hinder the adaptation of new sustainable shopping habits based on the amount of agents exposed to it and its inherent bias.\\

(B1) \fbox{$\delta = 0.1$, $\tau = \text{N/A}$, $\sigma = 0.1, 0.3, 0.5$, $\beta = 0$, $\zeta = \text{N/A}$}\\

We begin by isolating the effects of social media on a non-polarized agent set, assuming no bias and focusing on exposure levels. We find that medium exposure ($\sigma=0.3$) has the most positive effect on the changes in concerns (Fig. (\ref{fig:graphA4}), Appendix.) When exposure to social media is too low, not enough agents are influenced, leading to minimal change. Conversely, high exposure levels polarize agents due to the nature of the social media feedback loop function, increasing the number of anti-sustainability agents and reducing the average concern. This is consistent with previous findings on mass media effects on cultural dynamics (\cite{gonz_aacute_lez-avella2007}), which suggest that weak media influence promotes cultural homogeneity, while strong messaging sustains or enhances cultural diversity. This occurs because a significant portion of agents initially have high concerns, causing many to quickly polarize towards pro-sustainability. But  higher exposure impacts those who may become polarized in the opposite direction, increasing the number of anti-sustainability agents and hence bringing the average concerns down.\\

(B2)  \fbox{$\delta = 0.10$, $\tau = \text{N/A}$, $\sigma = 0.10$, $\beta = -0.30, -0.15, 0.00, 0.15, 0.30$. $\zeta = \text{N/A}$}\\

Next, we look at the different levels of biases on a non-polarized, ranging from very anti-sustainability ($\beta= -0.30$) to very pro-sustainability ($\beta=0.30$). We keep social media engagement constant (). We find that social media bias significantly impacts the effectiveness of communication in changing concerns. First, there is little difference in results between moderately pro-sustainability stances ($\sigma=0.15$ and $\sigma=0.30$), suggesting that even a moderate pro-sustainability bias can achieve a positive impact. However, moderately and strongly anti-sustainability stances ($\sigma=-0.15$ and $\sigma=-0.30$) hinder progress considerably (Fig. \ref{fig:graphA5}, Appendix). Social media often promotes increased consumption, especially in clothing, through in-platform shops, algorithms favoring shopping hauls, ads, and the ability to tag clothing in posts. This pro-consumption bias poses a significant threat to sustainable shopping habits, trapping individuals in a cycle of overconsumption. Conversely, if platforms favor sustainable content, they can promote more sustainable shopping habits (\cite{sm_sust1}, \cite{sm_sust2}, \cite{sm_sust3}). \\

(B3)  \fbox{$\delta = 0.40$, $\tau = \text{N/A}$, $\sigma = 0.40$, $\beta = -0.15, +0.15$, $\zeta = \text{N/A}$}\\

Moreover, the strong influence of social media persists in non-polarized agent sets with increased communication and social media exposure. In these cases of moderate influence, there are significant differences in overall changes, as shown in Fig. (\ref{fig:graph1}). The associated variances are illustrated in Fig. (\ref{fig:graph2_vars}) in the Appendix, where variability in opinions is significantly larger with pro-sustainability social media bias compared to pro-consumption bias. This might be due to the initially high average probability of buying fast fashion at the start of the simulation. From these simulations, we learn that increased communication between agents is unlikely to combat social media, and significant influence of platforms remains dominant.

\begin{figure}[ht]
\centering
\includegraphics[width=\textwidth]{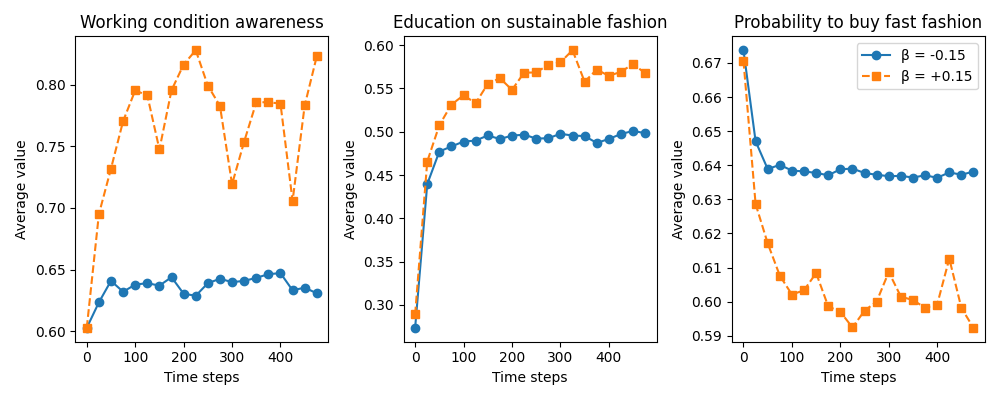}
\caption{Changes in average values for Model (B3)
\textit{Source: our simulations.}}
\label{fig:graph1}
\end{figure} 

(B4)  \fbox{$\delta = 0.4$, $\tau = 0.15$, $\sigma = 0.10, 0.30, 0.50$, $\beta = 0$, $\zeta = \text{N/A}$}\\

For the polarized agent set, we examined the same scenario of different levels of exposure to social media, but with initially increased communication. We compared different levels of social media engagement. We found that social media has a smaller impact on overall changes in concerns compared to the non-polarized population (Fig. \ref{fig:graphA6}, Appendix). Additionally, lower levels of engagement ($\sigma = 0.1$) positively impact concerns, which contrasts with the findings for the non-polarized agent set. On the contrary, medium ($\sigma = 0.3$) higher levels ($\sigma=0.5$) of engagement had similar effects and impacted concerns and probability to buy fast fashion to a smaller extent. This was not the case for the non polarized agentset, highlighting how the same influence can impact agents differently depending on the nature of their communication.\\

(B5) \fbox{$\delta = 0.1$, $\tau = 0.15$, $\sigma = 0.35$, $\beta = -0.30, -0.15, 0.00, 0.15, 0.30$}\\

\begin{figure}[ht]
\centering
\includegraphics[width=\textwidth]{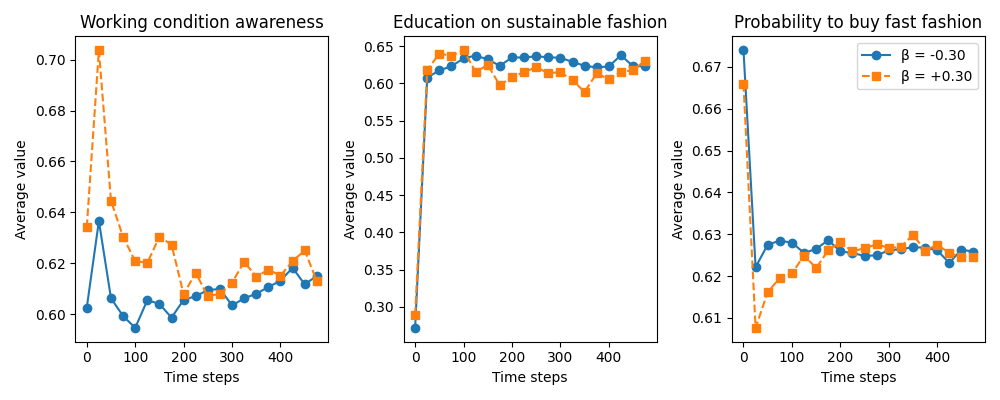}
\caption{Changes in average values for Model (B4)
\textit{Source: our simulations.}}
\label{fig:graph2}
\end{figure}

We observe a smaller impact of biases on the polarized agentset, compared to the non-polarized agent set, as shown in Fig. (\ref{fig:graph2}). This impact remains modest even with increased social media engagement. However, positive social media biases increase variability in all concerns and opinions, unlike negative biases, which show less impact on variability. By comparing Fig. (\ref{fig:graph1}) and Fig. (\ref{fig:graph2}), we see that social media biases affect the non-polarized agent set more significantly than the polarized agent set. This difference highlights how social media influences populations based on their preexisting views and the strength of their opinions.

Our findings reveal that social media significantly impacts concerns and opinions, with the extent of this impact varying based on the polarization of the agent set. For non-polarized agents, moderate pro-sustainability social media exposure ($\sigma=0.30$) enhances positive changes in concerns, while extreme biases towards pro-consumption or anti-sustainability can hinder progress. In contrast, for polarized agents, social media’s influence is less pronounced, and lower engagement levels have a more positive effect compared to higher levels. Additionally, biases have a smaller impact on polarized agents compared to non-polarized ones. Overall, our simulations indicate that social media’s effects are more pronounced in non-polarized populations and underscore how preexisting views and the nature of communication influence the effectiveness of social media in shaping opinions.

\subsection{Set C: government interventions}

A major goal of this study is to find the effect that campaigns and government sponsored advertisements can have on the population. The aim is to predict policy impacts and public opinion shifts to provide recommendations for policy making. \\

(C1)  \fbox{$\delta = 0.4$, $\tau =\text{N/A}$, $\sigma = 0.10$, $\beta = 0$, $\zeta = 0.5, 0.8, 1.0, 1.2, 1.5$}\\

\begin{figure}[ht]
\centering
\includegraphics[width=\textwidth]{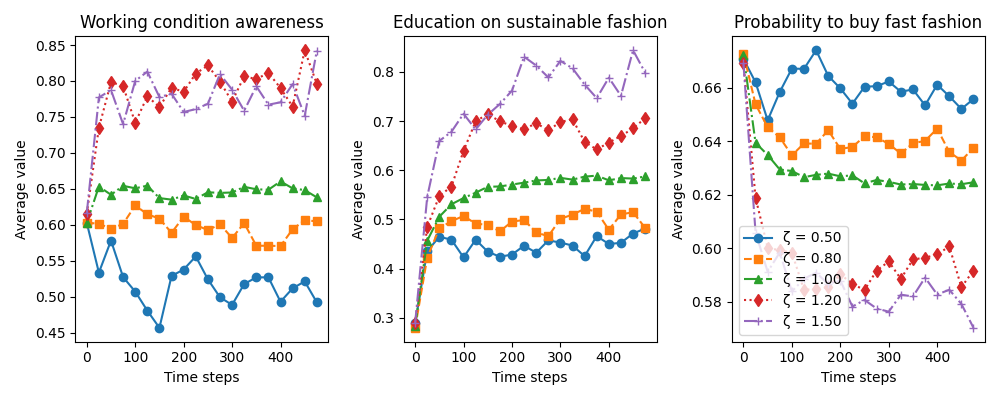}
\caption{Changes in average values for Model (C1)
\textit{Source: our simulations.}}
\label{fig:graphA7}
\end{figure}

\begin{figure}[ht]
\centering
\includegraphics[width=\textwidth]{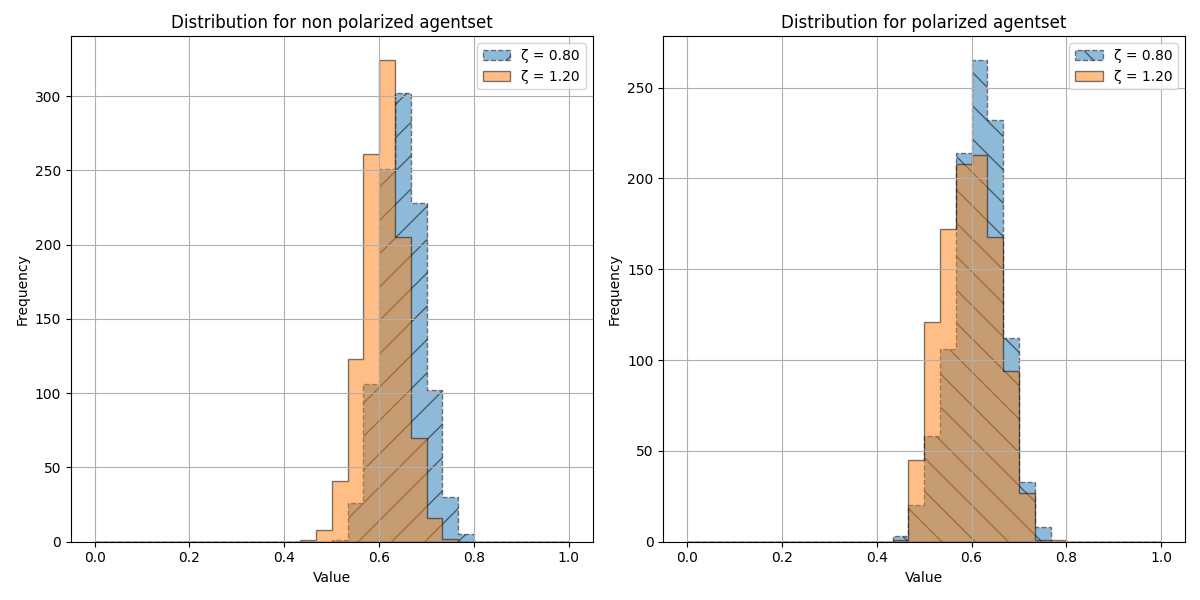}
\caption{Final distributions (left: C1, right: C2) to purchase from a fast fashion brand with pro-fast fashion (blue, crosshatch patterns) and pro-sustainability (orange, no patterns) campaigns.
\textit{Source: our simulations.}}
\label{fig:graph34}
\end{figure}

We start by examining how state interventions influence shifts in average opinions within non-polarized agentsets. In our model, the state’s stance, represented by the value of $\zeta$, ranges from $\zeta=0.5$ (anti-sustainability) to $\zeta=1.5$ (pro-sustainability), reflecting different overall goals. Our initial focus is on the effects of government stances with increased communication levels. This choice is based on research suggesting that once a topic becomes politically or religiously prominent, it is likely to become polarizing and provoke increased communication (\cite{politicization_polarization}). Consequently, we limit our study to scenarios with elevated peer influence. We observe that agents are significantly influenced by government campaigns and stances, as expected in a non-polarized context. Interestingly, there is little difference in the outcomes between moderately pro-sustainability ($\zeta=1.2$) and highly pro-sustainability ($\zeta=1.5$) campaigns, as seen in Fig. (\ref{fig:graphA7}). This suggests that a moderately pro-sustainability stance is sufficient for effective results, without needing an extreme position. Additionally, the probabilities of purchasing fast fashion cluster similarly for both moderately pro-fast fashion ($\zeta=0.8$) and pro-sustainability ($\zeta=1.2$) states, as shown by the individual distributions at the end of both simulations in the left histogram of Fig. (\ref{fig:graph34}). The clustering does occur around slightly different values, but the overall the shape of the distribution is very similar. 

(C2) \fbox{$\delta = 0.4$, $\tau = 0.15$, $\sigma = 0.10$, $\beta = 0$, $\zeta = 0.5, 0.8, 1.0, 1.2, 1.5$}\\

Next, we study the same scenarios for a polarized agentset. The results reveal notable differences compared to the non-polarized agent set. We observe greater resistance to anti-sustainability campaigns and smaller variability in outcomes across different government stances, indicating that polarized agents are less likely to change their opinions quickly. This resistance aligns with the idea that once a topic becomes politicized, it tends to be more controversial. Additionally, the final distributions of individual probabilities are slightly more heavy-tailed in the polarized agent set than in the non-polarized one, especially for the pro-sustainability campaigns. This can be seen in Fig. (\ref{fig:graph34}), where we plot the final distributions of individual probabilities for models with moderately pro-fast fashion ($\zeta=0.8$) and pro-sustainable ($\zeta=1.2)$ states.

(C3)  \fbox{$\delta = 0.4$, $\tau = 0.15$, $\sigma = 0.40$, $\beta = -0.30, 0.00, +0.30$, $\zeta = 1.2$}\\

The next question we explore is the role of social media in the policy-making process and its impact on progress. We previously noted that social media often exhibits pro-fast fashion biases, so we investigate how these and other stances influence shifts in shopping habits. We examine pro-fast fashion, neutral, and pro-sustainability social media biases within a polarized agent set, considering increased communication and social media engagement with a pro-sustainability state. Our simulations show that social media plays a significant role in shaping agents' shopping habits (Fig. (\ref{fig:graphA9}), Appendix.) It can either hinder or enhance progress, highlighting the importance for policy-makers to account for social media's influence on the population. \\

(C4)  \fbox{$\delta = 0.4$, $\tau = 0.10, 0.20, 0.30$, $\sigma = 0.10$, $\beta = 0$, $\zeta = 1.2$}\\

Realistically, campaigns cannot continue indefinitely due to financial constraints and the diminishing impact of repeated exposure, a phenomenon known as ``campaign fatigue" (\cite{campaign_fatigue}). To address this, we investigate the effects of halting moderate pro-sustainability campaigns ($\zeta=1.2$) mid-simulation (around 250 time steps). Our focus is on a polarized agent set with varying levels of polarization. Given that higher tolerance reduces perceived controversy, we explore how different levels of polarization affect campaign effectiveness, using three tolerance levels with increased communication levels. We find that greater tolerance (lower polarization, $\tau=0.30$) results in more enduring effects, while the lowest tolerance ($\tau=0.10$) leads to a significant decay in concerns, often returning to pre-campaign levels (Fig. (\ref{fig:graphA10}), Appendix.) In contrast, the non-polarized ($\tau=0.5$) agentset shows less promising results, where the achieved awareness levels fall to pre-campaign levels. Our findings suggest that a slightly polarized agent set is most likely to sustain the effects of a campaign over time.\\

(C5) \fbox{$\delta = 0.4$, $\tau = 0.15$, $\sigma = 0.10$, $\beta = -0.30, +0.30$, $\zeta = 1.2$}\\

\begin{figure}[ht]
\centering
\includegraphics[width=\textwidth]{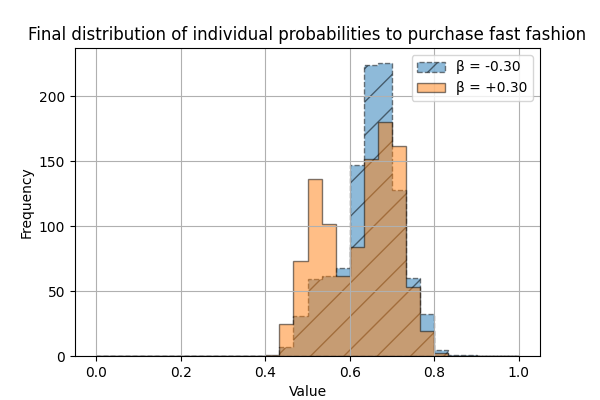}
\caption{Final distributions (C5) to purchase from a fast fashion brand and a halt of pro-sustainability campaigns with pro-fast fashion (blue, crosshatch patterns) and anti-fast-fashion (orange, no patterns).
\textit{Source: our simulations.}}
\label{fig:graph6}
\end{figure}

\begin{figure}[ht]
\centering
\includegraphics[width=\textwidth]{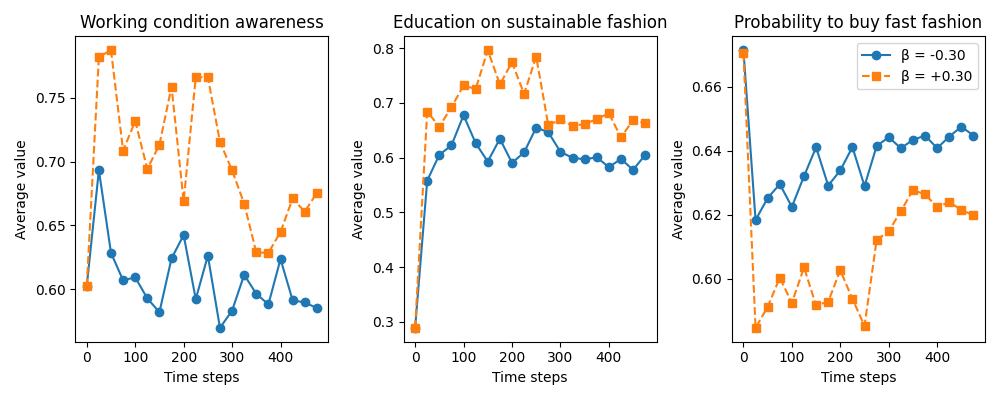}
\caption{Changes in average values for Model (C5)
\textit{Source: our simulations.}}
\label{fig:graph5}
\end{figure}

In light of this, we also examine how social media biases impact the effects of halting campaigns. We model two scenarios: one with a pro-fast fashion bias and another with a pro-sustainability bias. From Fig. (\ref{fig:graph5}), we can see that social media has a significant impact both during and after the campaigns. Additionally, the agents' probabilities of purchasing fast fashion differ qualitatively by the end of each simulation. Fig. (\ref{fig:graph6}) illustrates that with a pro-sustainability social media bias, there is greater variance in the final probabilities, whereas with a pro-fast fashion bias, the probabilities are more clustered around a single value.

Our investigation into government interventions reveals that the effectiveness of campaigns varies significantly between non-polarized and polarized agent sets. For non-polarized agents, moderate pro-sustainability campaigns lead to substantial shifts in opinions, with only minor differences between moderately and highly pro-sustainability stances. In polarized agent sets, the resistance to anti-sustainability campaigns is greater, and the variability in outcomes is reduced, reflecting a slower rate of opinion change. When campaigns are halted, higher tolerance levels (less polarization) result in more enduring effects, while low tolerance levels (high polarization) lead to a sharp decline in concerns. Social media biases further complicate this landscape: pro-sustainability biases increase variability in opinions, while pro-fast fashion biases lead to more clustered, less varied outcomes. Overall, the simulations suggest that while government campaigns are influential, their impact is mediated by the level of polarization and the prevailing social media biases. Most importantly, government interventions appear to set the overall tone for progress, and the changes they initiate are not observed to be achievable without the state's campaigns.

\section{Conclusion} \label{conclusion}
In this study, we examined the relationship between people's concerns and their purchasing behavior. The data used in this study indicates that environmental concerns do not significantly influence purchasing habits, nor do normative expectations. This suggests a general unawareness of the environmental impact of fast fashion and a lack of societal pressure to shop sustainably. The unawareness may be a result of insufficient education and could be addressed through targeted educational efforts. However, the absence of societal pressure is more troubling. Societal pressure is crucial for initiating discussions and shaping collective behavior. Its absence suggests that even those educated on the issue may choose to remain silent. In this study, we examined what happens when normative expectations are introduced artificially, and how its impacts couple with social media influence and government efforts to educate the population.

Our findings from  models with increased socialization reinforce previous research findings on the importance of opinion sharing for fostering acceptance and supporting public conversations on challenging topics (\cite{communication2}, \cite{communication3}). In particular, large-scale shifts in public opinion depend considerably on the overall receptiveness of the broader population (\cite{receptiveness}). This underscores the importance of cultivating environments where individuals are open to influence and engaged in peer-to-peer discussions, as collective change is more likely to emerge when many moderately connected people are socially primed to update their views.

We find that social media profoundly influences concerns and opinions, with the degree of its impact depending on the polarization level of the agent set. Additionally, social media overall bias has a great impact on efforts to change current fashion purchasing habits. We found that it can both help progress and hinder the adaptation new shopping habits, hence playing a powerful role in shaping our behavior.

One goal of this study was to establish sensible measures for government implementation based on their objectives and the characteristics of the society in question. One of the key findings is that the government does not need to adopt an extremely proactive stance to achieve optimal results. We discovered that the state's influence on public opinion reaches a point where further interventions yield insignificant returns. This indicates that more interventions are not necessarily better, since there is a limit to how much the state can affect its population. These results are consistent with existing studies that explore mechanisms to overcome public inaction and drive widespread pro-environmental behavior change (\cite{abm_environment_2}).

Most importantly, our study emphasizes the state's crucial role in initiating these conversations and the readiness of individuals to engage with them. Relying solely on social media or highly concerned individuals to introduce these topics is not effective and tends to stagnate the process.

While our study provides valuable insights, it is based on several simplifying assumptions that could be refined in future research. For example, agents currently form friendships randomly and do not have the ability to end or create new ties based on opinion differences. Allowing such dynamics could lead to more realistic formations of homogeneous groups, reflecting real-world social behavior more closely. Similarly, interactions are modeled as unilateral and equally influential for all agents, whereas in reality, interactions tend to be bilateral with varying levels of influence among individuals. Additionally, agents interact on only one topic per time step, which complicates the direct correspondence between model time and real-world timelines. This relates to another limitation our study faces, which is the lack of longitudinal data necessary for the validation of the model. 

Regarding government interventions, the model assumes that the government has access to the average opinion of the entire population, which is unlikely in practice. Moreover, agent susceptibility to government actions is static, though in reality, individuals’ attitudes might change dynamically depending on whether they support or oppose the state’s policies. In terms of social media influence, the model excludes a small segment of non-users—about 6.8$\%$ of the relevant population—which makes this assumption reasonable. However, susceptibility to social media is held constant throughout the simulation, despite evidence that susceptibility tends to increase with usage. The model also assumes that social media acts as a polarizing force, although empirical findings on this are mixed. Additionally, the three types of susceptibility are assigned independently of one another. Exploring the results when these are interconnected can be of interest, although at this time there is no evidence to support the existence of such dependence. 

Finally, the decision-making model is specifically calibrated to the Spanish market, which may limit the applicability of our results to other populations. Cultural differences and varying values placed on clothing both across and within countries mean that the model would need to be adjusted using local data to be relevant elsewhere.

Addressing the overconsumption of garments requires a cultural shift in post-industrial countries. People must recognize that the right to be fashionable should not outweigh the inalienable right of others not to be oppressed and exploited. Those who understand and care about these issues, who have the social support, resources, psychological health, and the freedom to explore and to innovate, must step up. The responsibility for change cannot rest solely on those who are systemically and systematically oppressed, and at risk of suffering from environmental consequences; it must also come from those who are in the position to perpetuate the current system, but choose to do otherwise. No one is inherently entitled to be fashionable, just as no one is destined to be exploited.

\section*{Acknowledgment}
We would like to thank Dr. Silvia Blas Riesgo, for generously sharing her database on Spanish consumers. 

This work was supported by the Ministry of Education, Vocational Training and Sports (MEFPD).

A.S. acknowledges support from project PID2022-141802NB-I00 (BASIC) funded by MCIN/AEI/10.13039/501100011033 and by “ERDF A way of making Europe”, and from grant MapCDPerNets---Programa Fundamentos de la Fundación BBVA 2022.

\section*{Declaration of generative AI and AI-assisted technologies in the writing process}
During the preparation of this work the author(s) used ChatGPT 3.5, ChatGPT 4.0, and Elicit.com in order to rewrite of paragraphs previously written by the authors and find bibliography and references for searched topics.After using this tool/service, the author(s) reviewed and edited the content as needed and take(s) full responsibility for the content of the published article.

\printbibliography

\smallskip
\newpage

\appendix
\section{Appendix}
\renewcommand\thefigure{\thesection.\arabic{figure}}    
\setcounter{figure}{0} 
\subsection{Decision-making model.} \label{decision_making_appendix}

In this segment, we look into the exogenous variables and their calculation for each agent. The omitted variables in this analysis are outlined in Section. Participants were prompted to assess their agreement level with statements on a scale from 1 (strongly disagree) to 5 (strongly agree). Statements marked with an asterisk (*) were reversed for regression analysis.

\begin{enumerate}
    \item \textbf{Environmental concerns:} this variable denotes the extent of environmental concern for each agent. It was derived from individuals' ratings of the following statements:
    \begin{itemize}
        \item I am worried about the environment.
        \item The conditions of the environment influence the quality of my life.
        \item I think it is important to protect and preserve the Earth for future generations.
        \item I think that the environmental crisis is being exaggerated*.
        \item I believe sustainability is important.
    \end{itemize}

    \item \textbf{Working conditions awareness }: this variable reflects an individual's concern for the working conditions of people in the garment industry. It was determined by the ratings people gave to the following statements:
     \begin{itemize}
        \item The working conditions are something I worry about when buying clothing or accessories.
        \item When I buy clothing or accessories, I take into consideration whether they have been produced under fair trade practices.
        \item That workers receive a fair salary is important to me when buying clothes or accessories.
    \end{itemize}

    \item \textbf{Shopping frequency:} this variable captures the frequency of people's shopping activities, including both online and offline purchases, during and outside of sales periods. Participants rated the following statements on a scale ranging from 1 (never/almost never) to 6 (multiple times a week):

    \begin{itemize}
        \item How often do you buy clothes or accessories? [Offline].
        \item How often do you buy clothes or accessories? [Online].
        \item How often do you buy clothes or accessories during sales? [Offline].
        \item How often do you buy clothes or accessories during sales? [Online].
    \end{itemize}

    \item \textbf{Education level:} this variable accounts for individuals' knowledge about sustainable fashion. It was computed from participants' dichotomous (Yes/Agree or No/Disagree) responses to the following questions/statements:
    \begin{itemize}
        \item Have you ever heard of sustainable fashion?
        \item Can you define sustainable fashion?
        \item Can you name any sustainable fashion brands?
        \item I do not know what sustainable fashion is.
    \end{itemize}

    \item \textbf{Normative expectations:} the importance of what we perceive others expect from us significantly influences our decision-making process. We look at individuals' agreement levels with the following statement to understand their normative expectations regarding sustainable shopping:
    \begin{itemize}
        \item My family and friends expect me to buy more sustainable products.
    \end{itemize}

    \item \textbf{Trust:} individuals tended to purchase less sustainable products if they concurred with the following statement:
    \begin{itemize}
        \item Sometimes I'm not sure if a brand is truly sustainable or if it's just claiming to be to improve its image.
    \end{itemize}

    \item \textbf{No access to sustainable fashion}: 
    individuals who believed they lacked physical access to sustainable brands tended to consume more fast fashion products.

    \item Demographic variables such as gender, age, income, and level of education were also incorporated into the multilinear regression analysis.
    
\end{enumerate}

The response variable we aimed to explain using the predictor variables is the likelihood of engaging in fast fashion consumption. This metric was computed for each individual based on their responses and level of agreement with the following questions/statements:

\begin{itemize}
    \item Choose ONE option:\\
         1) I mainly buy clothes when I need them, when something is damaged, or when it no longer fits me.\\
         2) I buy clothes that fit my personal style; I don't mind whether they are trendy or not.\\
         3) I go shopping because I like to stay up-to-date and follow trends.  
    \item In which stores or from which brands do you most frequently make purchases? (text input)
    \item Have you purchased any products from a sustainable fashion brand? (Yes/No)
    \item Have you ever bought second-hand clothing/accessories? (Yes/No)
    \item I often buy sustainable fashion. (Agree/Disagree)
    \item I prefer to buy second-hand clothing. (Agree/Disagree)
\end{itemize}

Several other variables were considered in the multilinear regression analysis but did not demonstrate statistical significance. Among the predictor variables that were discarded were:

\begin{enumerate}
    \item \textbf{Perceived Consumer Effectiveness:} according to the Value-Belief-Norm (VBN) theory of environmental behavior (\cite{VBN}), perceived consumer effectiveness influences purchasing power. However, our empirical data revealed no correlation between perceived consumer effectiveness regarding the impacts of shopping sustainably and individuals' likelihood of purchasing it. Interestingly, nearly everyone expressed high levels of agreement with the following statements:
    \begin{itemize}
        \item The behavior of each consumer can have a positive impact on society.
        \item I believe that I can generate a positive impact on the environment by consuming sustainable products instead of non-sustainable ones.
        \item I believe that buying sustainable clothing can help combat environmental issues.
    \end{itemize}
    Even though most individuals agreed with these statements, it did not predict whether they shop sustainably or not. The study (\cite{TPB_perceived_control}) also suggests that ``sustainable awareness" is not strongly associated with sustainable purchasing behavior. Since definition of ``sustainable awareness" combined our definitions of ``Environmental concerns" and ``Working conditions concerns," it is not straightforward whether our findings agree, but we also find that environmental concerns have little impact on people's purchasing behavior.

    \item \textbf{Normative expectations:} this variable is currently part of the multilinear regression model. Although we discovered a small correlation between individuals' behavior and their level of agreement with the statement "My family and friends expect me to buy more sustainable products.," there was no correlation found between the level of agreement with the following two statements and the likelihood of purchasing fast fashion:
    \begin{itemize}
        \item Society expects me to buy more sustainable products.
        \item I think I have a moral obligation to buy clothes/accessories made sustainably.
    \end{itemize}

    From this, we can conclude that individuals surrounded by other sustainability-conscious individuals might shop more sustainably. However, people's level of agreement with the two statements above had no impact on their behavior. This is a crucial finding because it indicates that societal pressure is not a significant factor shaping people's shopping behavior.
    
    \item \textbf{Income:} Interestingly, income played no role in people's shopping habits. Furthermore, it did not predict shopping frequency or the average amount individuals spend on clothes. A linear regression analysis revealed that income only accounted for $1.9\%$ of the variation in monthly spending on clothing, and similar results were found for shopping frequency. This can be attributed to the nature of the fashion industry, particularly the fast fashion segment. The consistent decline in garment prices, despite inflation, has been a historical trend (\cite{fashion_prices_drop}). Additionally, people now allocate a smaller portion of their income to clothing than in the past (\cite{spend_less_on_clothing_now}). While in the past, individual income significantly influences household expenditures on clothing (\cite{people_budget_depending_on_income_2}), and some recent studies suggest this continues to be the case (\cite{people_budget_depending_on_income}), it doesn't seem to hold true for the Spanish consumer market, as there was no correlation between income and monthly expenditure on clothing.

    \item \textbf{Education:} the level of education an individual had received had no effect on their clothing consumption behavior.

\end{enumerate}

\subsection{Agent-Based Model: additional details.}\label{agent_based_details_appendix}

\subsubsection{Polarized agentset - peer interaction function}

The equation used for social interactions in a polarized society will be similar to that for a non-polarized society, given by Eq. (\ref{peer-non-pol-eq}). However, the difference lies in the introduction of a tolerance threshold $\tau$, which can be set at the beginning of the simulation. This threshold ranges from $\tau=0.05$ to $\tau=0.5$ and represents how tolerant agents are toward opinions that deviate from theirs by more or less than the threshold.

If the threshold is set to $\tau=m$, then opinions within a range of $\pm m$ from an agent's opinion will have a homogenizing effect. However, opinions beyond this threshold will have an opposite effect, pushing the agent's opinion toward the opposite end of the spectrum. For instance, if an agent's opinion is $0.5$, with a tolerance level of $\tau=0.2$, then the agent will ``accept" an opinion within the range $[0.3, 0.7]$, but anything that falls outside of that opinion will have the opposite effect. In particular, it will make incentivize the agent to adopt beliefs opposite to that of the agent they are interacting with. Since opinions are defined in the range $[0,1]$, for a given opinion $b\in[0,1]$, we define the opposite opinion as $1-b$ such. Thus, if an agent $i$ with opinion $b_i=0.3$ interacts with an agent $j$ with opinion $b_j=0.9$ while the tolerance threshold is $\tau=0.2$, then the $j$th agent's opinion will contribute $1-b_j=0.1$ to $i$th agent's opinion on the topic. The role of $\tau$ is similar to the role of Broadening Behavior introduced in (\cite{broadening_behavior}), in a sense that it introduces a kind of repulsion towards the opinion of others, especially those an agent disagrees with, and encourages agents to diverge. However, it is important to note that the tolerance threshold we introduce is different from the broadening behavior since it does not encourage the agent to be different from their neighbors, but rather to differ more from the group the agent already differs from. 

\subsubsection{Social media and government feedback loops}

\begin{figure}[ht]
     \centering
     \begin{subfigure}[b]{0.3\textwidth}
         \centering
         \includegraphics[width=\textwidth]{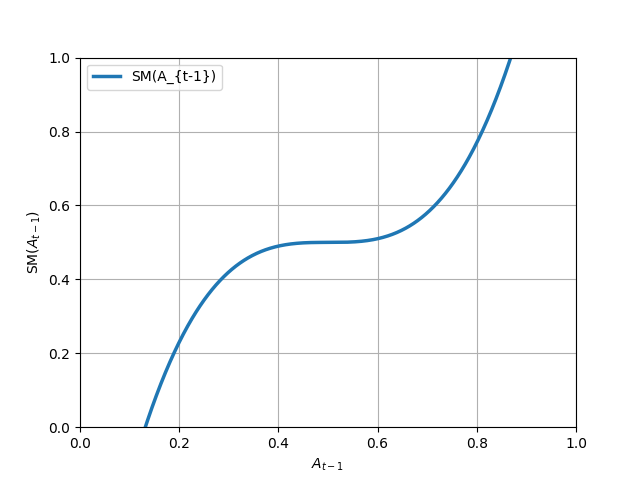}
         \caption{$S_{sm}=0.2$}
         \label{fig:A1a}
     \end{subfigure}
     \hfill
     \begin{subfigure}[b]{0.3\textwidth}
         \centering
         \includegraphics[width=\textwidth]{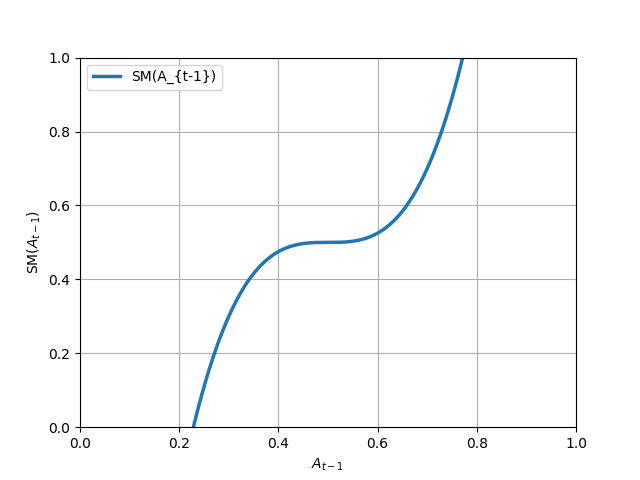}
         \caption{$S_{sm}=0.5$}
         \label{fig:A1b}
     \end{subfigure}
     \hfill
     \begin{subfigure}[b]{0.3\textwidth}
         \centering
         \includegraphics[width=\textwidth]{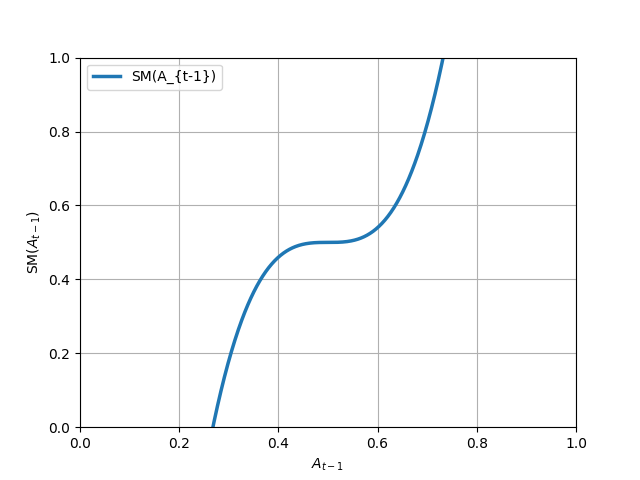}
         \caption{$S_{sm}=0.8$}
         \label{fig:A1c}
     \end{subfigure}
        \caption{Social media feedback loop function from Eq. (\ref{feedback_sm_equation}) for $\beta=0$ and different $S_{sm}$ values.}
        \label{fig:A1}
\end{figure}

Our feedback loop equation for social media, presented in Eq. (\ref{feedback_sm_equation}), is designed to simulate real-world algorithms that customize content for users based on their past preferences. This function also incorporates each agent's susceptibility. Fig.(\ref{fig:A1}) illustrates three variations of the social media feedback loop function, each reflecting different levels of susceptibility. In Fig. (\ref{fig:A1a}), the susceptibility is set at $S_{sm}=0.2$, indicating a low susceptibility, which results in a wider function. This means the content shown to the agent is similar to what they currently view, with minimal deviation. Fig. (\ref{fig:A1b}) depicts a susceptibility of $S_{sm}=0.5$, leading to a steeper function and a greater divergence from the agent's existing content. Finally, Fig. (\ref{fig:A1c}) shows a susceptibility of $S_{sm}=0.8$, resulting in the steepest function of the three, meaning the content presented is significantly more extreme compared to what the agent's current view. In any case, if the output exceeds $1$, it is manually capped at $0.95$ to prevent extreme values. Similarly, if the output is negative, it is adjusted to $0.05$ to avoid values outside the defined range.

\begin{figure}[ht]
     \centering
     \begin{subfigure}[b]{0.3\textwidth}
         \centering
         \includegraphics[width=\textwidth]{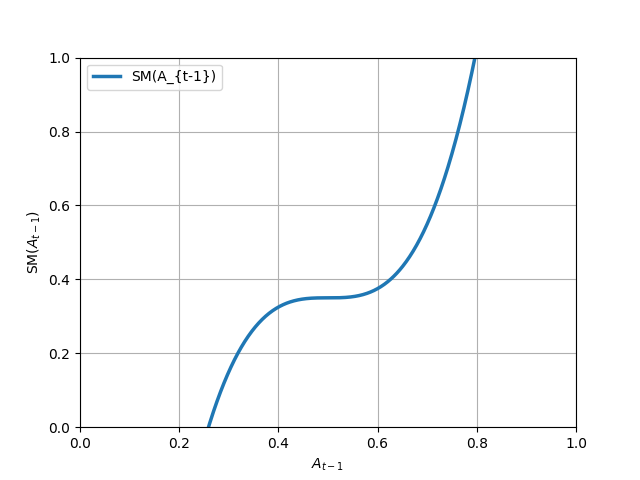}
         \caption{$\beta=-0.15$}
         \label{fig:A2a}
     \end{subfigure}
     \hfill
     \begin{subfigure}[b]{0.3\textwidth}
         \centering
         \includegraphics[width=\textwidth]{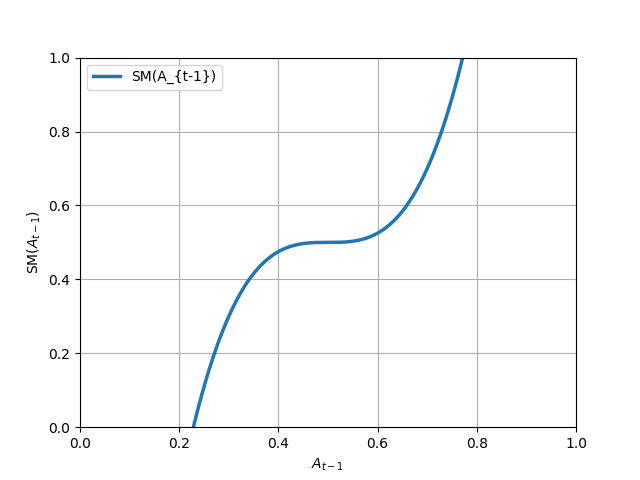}
         \caption{$\beta=0.00$}
         \label{fig:A2b}
     \end{subfigure}
     \hfill
     \begin{subfigure}[b]{0.3\textwidth}
         \centering
         \includegraphics[width=\textwidth]{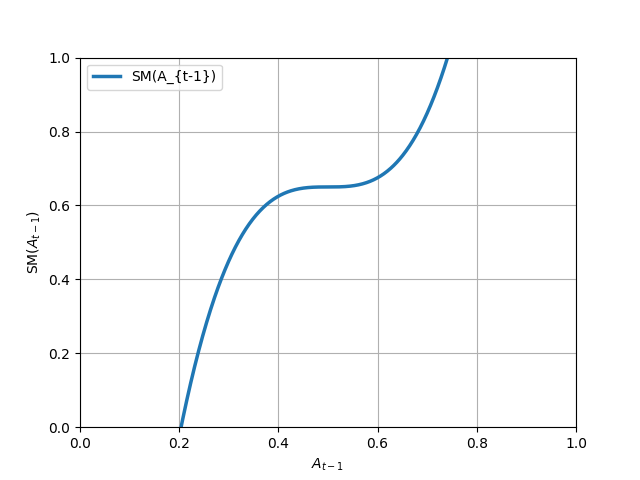}
         \caption{$\beta=+0.15$}
         \label{fig:A2c}
     \end{subfigure}
        \caption{Social media feedback loop function from Eq. (\ref{feedback_sm_equation}) for $S_{sm}=0.5$ and different bias values.}
        \label{fig:A2}
\end{figure}

Moreover, the function differs based on the value of $\beta$, and this adjustment is made for all agents. In Fig. (\ref{fig:A2}), we see the same feedback loop function for $S_{sm}=0.5$, but different levels of social media biases: anti-sustainability in Fig (\ref{fig:A2a}), neutral in Fig. (\ref{fig:A2b}), and pro-sustainability in Fig. (\ref{fig:A2a}).

Next, Eq. (\ref{social_media_influence}) ensures that an agent's opinion changes gradually and cannot be drastically affected in a single time step. Let us delve deeper into Eq. (\ref{social_media_influence}). Typically, a linear equation is used to adjust an agent's opinion based on an influence, and we plot an example of such in all upcoming figures for comparison. However, we devised a more sophisticated equation to ensure that an agent's opinion cannot change drastically in just one step. In Eq.(\ref{social_media_influence}), we make use of function $f$ and $g$, which are defined in Eq. (\ref{f}) and Eq. (\ref{g}), respectively.

\begin{equation}
    f(A_{t-1}, SM) = e^{-(1-S_{sm})\cdot |A_{t-1} - SM|}
    \label{f}
\end{equation}
\begin{equation}
    g(A_{t-1}, SM) = e^{S_{sm}\cdot |A_{t-1} - SM|}
    \label{g}
\end{equation}

\begin{figure}[ht]
     \centering
     \begin{subfigure}[b]{0.3\textwidth}
         \centering
         \includegraphics[width=\textwidth]{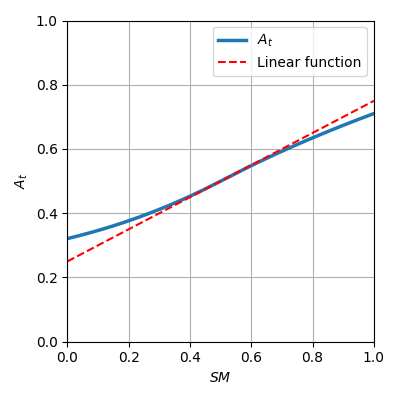}
         \caption{$A_{t-1}=0.5$}
         \label{fig:A3a}
     \end{subfigure}
     \hfill
     \begin{subfigure}[b]{0.3\textwidth}
         \centering
         \includegraphics[width=\textwidth]{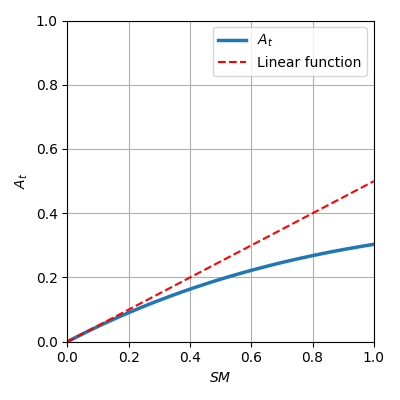}
         \caption{$A_{t-1}=0$}
         \label{fig:A3b}
     \end{subfigure}
     \hfill
     \begin{subfigure}[b]{0.3\textwidth}
         \centering
         \includegraphics[width=\textwidth]{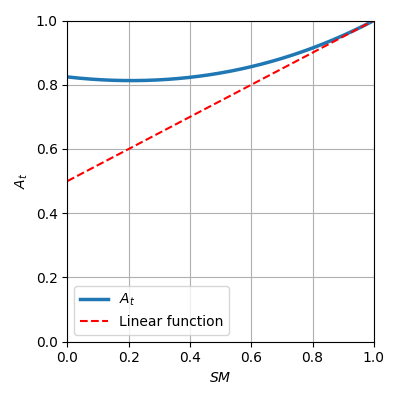}
         \caption{$A_{t-1}=1$}
         \label{fig:A3c}
     \end{subfigure}
        \caption{Social media influence function Eq. (\ref{social_media_influence}) for $S_{sm}=0.5$ and different inputs of feedback}
        \label{fig:A3}
\end{figure}

Consider an agent with a susceptibility to social media of $0.5$. The two inputs that are changing are the agent's current opinion and the social media ``influence" direction, calculated one step prior using the function described in Eq. (\ref{feedback_sm_equation}). Therefore, we have two inputs. To represent this in a 2D graph, we set the agent's opinion as a parameter, set at $A_{t-1}=0.0,0.5, 1.0$ and the x-axis represents the output of the social media function, indicating the opinion it promotes. The y-axis represents the agent's updated opinion. Fig. (\ref{fig:A3}) illustrates three functions for different values of $A_{t-1}$, and all figures contain a linear function, which is commonly used in such scenarios. Our function of choice assures that agent's opinion cannot be changed drastically in one time step, which cannot be guaranteed using a linear function.

Let us begin by examining Fig. (\ref{fig:A3a}), where the agent's opinion set at 0.5. We can see that, no matter what opinion is promoted by social media, agent's opinion will not change as drastically as in case of the red line. Next, let us look at Fig. (\ref{fig:A3b}), where agent's current opinion is at $A_{t-1}=0$. Note that with the linear function, the agent's opinion can change from 0 to 0.4 in just one time step, which is evidently too drastic. However, with the blue function, this is not the case, ensuring a smoother and slower transition. The further away from the original opinion - the smaller the derivative value of the blue line. Additionally, this implies that agents whose opinion correspond with what they see online are more receptive to the content, accounting for a common phenomenon observed in real life - confirmation bias.

Similarly, Fig. (\ref{fig:A3c}) illustrates the case where the agent's opinion is set at $A_{t-1}=1$. Here, the blue line depicts the potential return, while the red line serves as a reference. In all three cases, the blue line represents the values that the agent's opinion will take based on the input of social media influence. Notice that the derivative of the blue function decreases the further away we get from the intersection point, representing the smoother transition of opinion.

\begin{figure}[ht]
     \centering
     \begin{subfigure}[b]{0.45\textwidth}
         \centering
         \includegraphics[width=\textwidth]{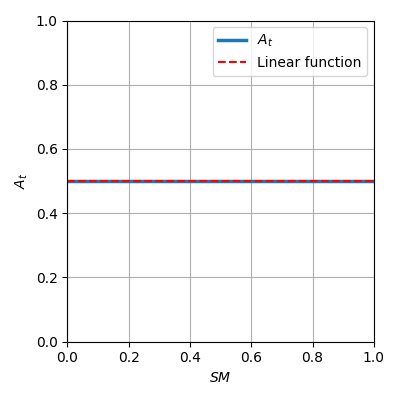}
         \caption{$S_{sm}=0.00$}
         \label{fig:A4a}
     \end{subfigure}
     \hfill
     \begin{subfigure}[b]{0.45\textwidth}
         \centering
         \includegraphics[width=\textwidth]{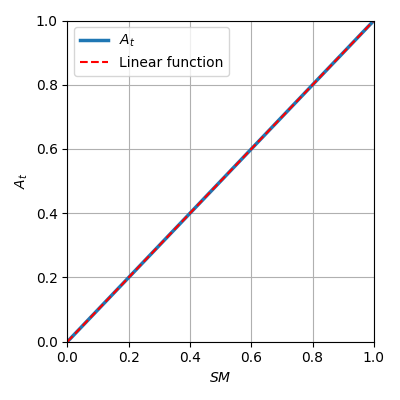}
         \caption{$S_{sm}=1.00$}
         \label{fig:A4b}
     \end{subfigure}
        \caption{Social media influence function Eq. (\ref{social_media_influence}) for different levels of susceptibility.}
        \label{fig:A4}
\end{figure}

Now, let's observe how this function changes as the susceptibility level varies. In Fig. (\ref{fig:A4}), we plot two cases: Fig. (\ref{fig:A4a}), where the agent's susceptibility is $S_{sm}=0.0$, and Fig. (\ref{fig:A4b}), where it is is $S_{sm}=1.0$. When the susceptibility level is zero, the input becomes irrelevant; social media will not influence the agent's opinion, thus returning the same value as $A_{t-1}$, which acts as a parameter for the plot. Similarly, for a susceptibility of 1, the function becomes a linear equation, as the agent is willing to undergo drastic opinion changes in a single step.

\subsection{Simulations} \label{simulations_appendix}

\subsubsection{Time management} 

At every time step, also referred to as ``tick," three events occur, managed by an interaction management function: one interaction with a specific number of friends, one instance of using social media, and one potential interaction with government campaigns. While the first two events occur deterministically once every time step, the occurrence of government campaigns is randomized and may or may not take place in an agent's day.

Furthermore, social media interaction occurs only once per day for simplicity, but it yields qualitatively satisfactory results. Increased usage of social media by some individuals is factored into the calculations through susceptibility towards social media: the more an individual uses social media, the more susceptible they are to its influence.

\subsubsection{Simulation}

Once the simulation starts, the main loop begins running until a set limit of ticks is reached, which can be controlled by the user and is set to 500 ticks in all the models presented in this thesis. The main purpose of this loop is to call the interaction management function and update the visual for the user. The interaction management function consists of initiating one of the three interactions of the time step for the agent. At the end of each time step, each agent's probability to purchase fast fashion is evaluated given the updates in their views and habits.

\subsubsection{Pseudocode}

We include a pseudocode to represent in a simplistic manner what the overall structure of the code is. In the next sections, we delve into each of the steps involved in it and justify their settings.

\begin{tcolorbox}[colback=white,colframe=gray,width=\linewidth, title= Overview of the model structure]
\centering

\begin{enumerate}
    \item Initialize agents and build visualization (setup).
    \item Go (for 500 ticks):
    \begin{enumerate}
        \item Random campaign is selected.
        \item For each agent:
        \begin{enumerate}
            \item Peer interaction.
            \item Social media interaction.
            \item Maybe: interaction with state's campaign.
        \end{enumerate}
        \item Update individual attributes based on changes.
    \end{enumerate}
\end{enumerate}

\end{tcolorbox}

\subsection{Figures} 

\begin{figure}[ht]
\centering
\includegraphics[width=0.85\textwidth]{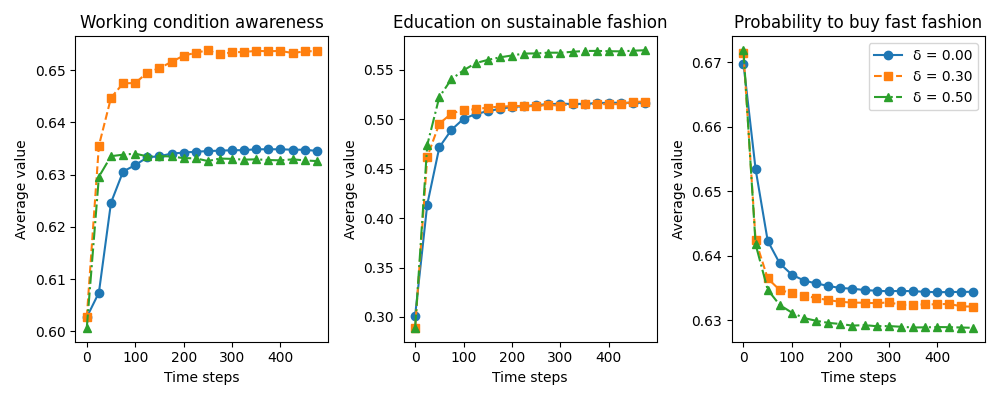}
\caption{Changes in average values for Model (A1)
\textit{Source: our simulations.}}
\label{fig:graphA1}
\end{figure}

\begin{figure}[ht]
\centering
\includegraphics[width=0.85\textwidth]{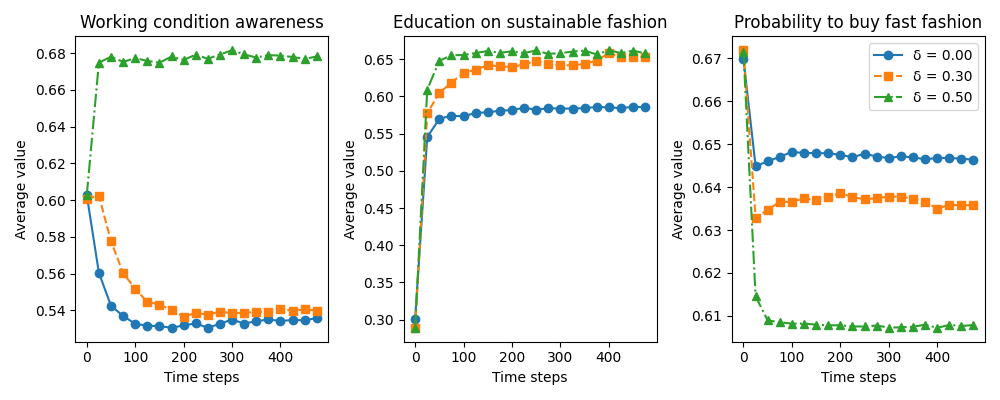}
\caption{Changes in average values for Model (A2)
\textit{Source: our simulations.}}
\label{fig:graphA2}
\end{figure}

\begin{figure}[ht]
\centering
\includegraphics[width=0.85\textwidth]{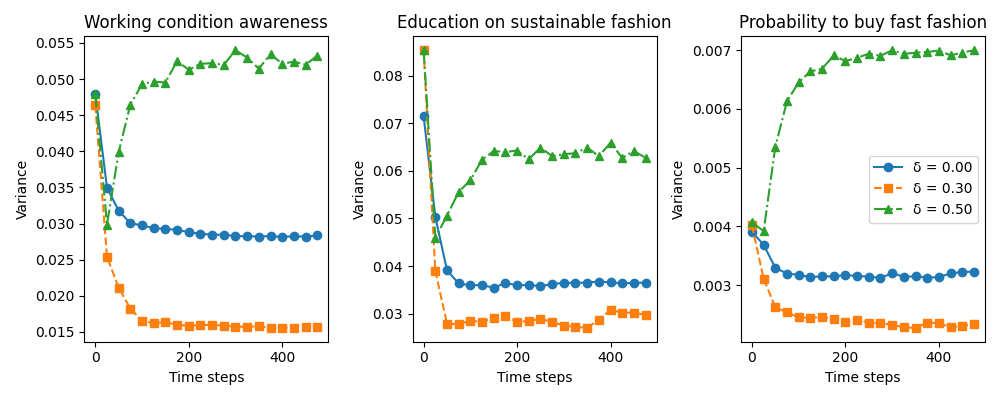}
\caption{Variances of average values for Model (A2)
\textit{Source: our simulations.}}
\label{fig:graphA2_vars}
\end{figure}

\begin{figure}[ht]
\centering
\includegraphics[width=0.9\textwidth]{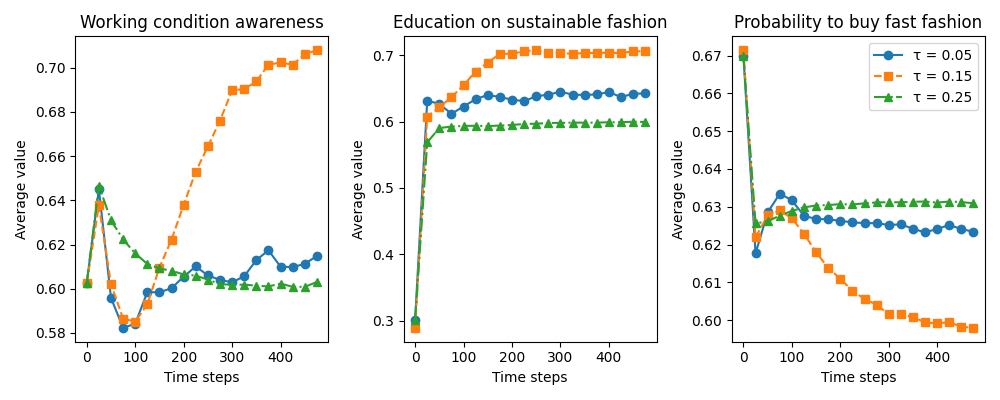}
\caption{Changes in average values for Model (A3)
\textit{Source: our simulations.}}
\label{fig:graphA3}
\end{figure}

\begin{figure}[ht]
\centering
\includegraphics[width=0.9\textwidth]{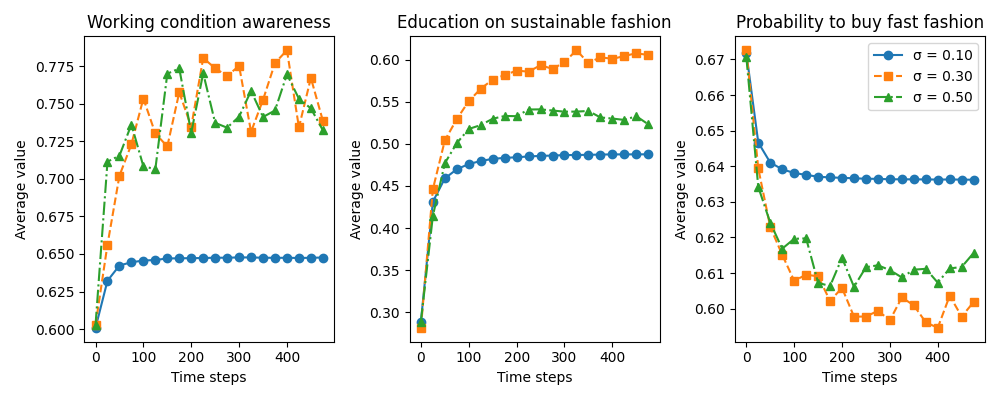}
\caption{Changes in average values for Model (B1)
\textit{Source: our simulations.}}
\label{fig:graphA4}
\end{figure}

\begin{figure}[ht]
\centering
\includegraphics[width=0.9\textwidth]{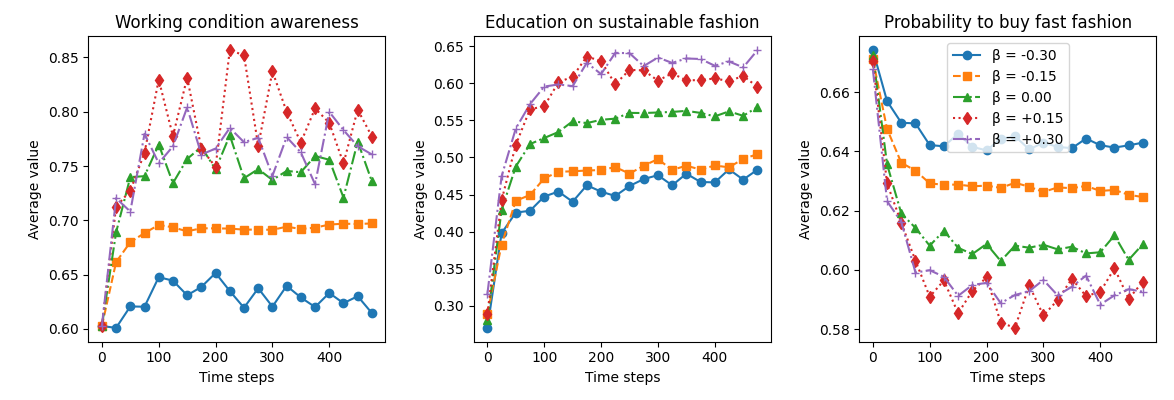}
\caption{Changes in average values for Model (B2)
\textit{Source: our simulations.}}
\label{fig:graphA5}
\end{figure}

\begin{figure}[ht]
\centering
\includegraphics[width=0.9\textwidth]{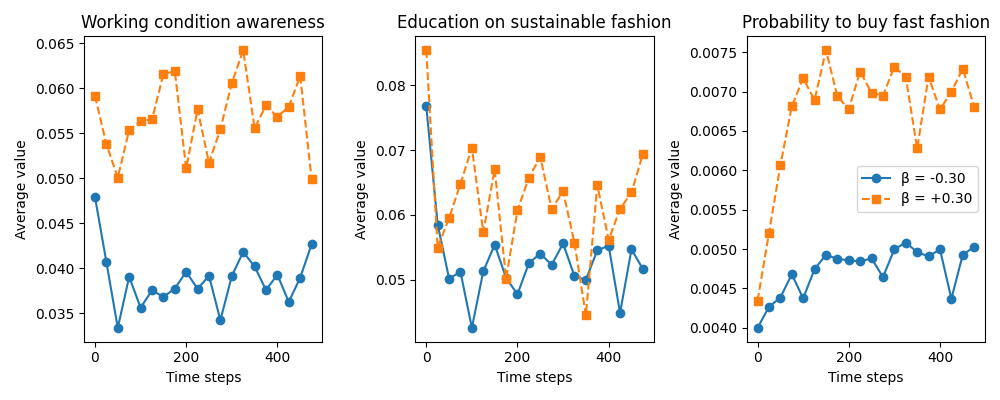}
\caption{Variances of average values for Model (B3)
\textit{Source: our simulations.}}
\label{fig:graph2_vars}
\end{figure}

\begin{figure}[ht]
\centering
\includegraphics[width=0.9\textwidth]{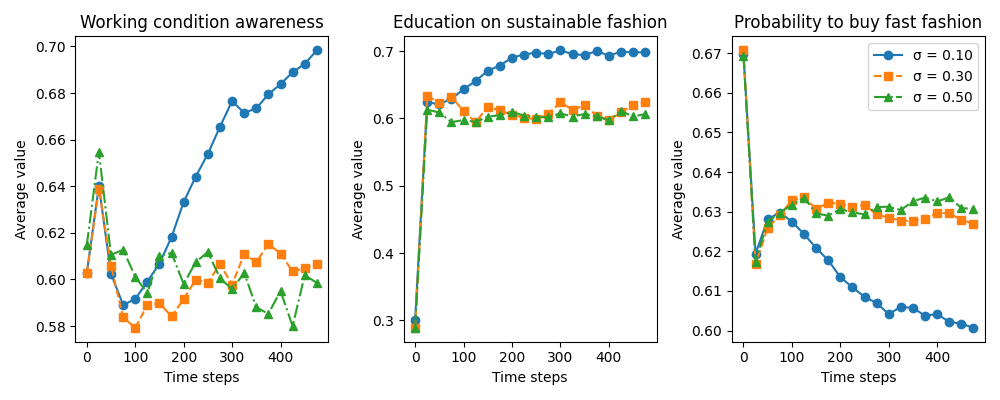}
\caption{Changes in average values for Model (B4)
\textit{Source: our simulations.}}
\label{fig:graphA6}
\end{figure}

\begin{figure}[ht]
\centering
\includegraphics[width=0.9\textwidth]{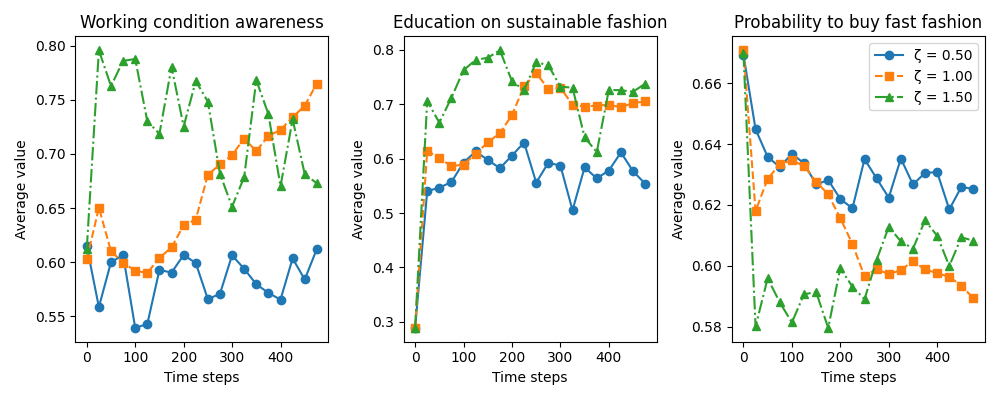}
\caption{Changes in average values for Model (C2)
\textit{Source: our simulations.}}
\label{fig:graphA8}
\end{figure}

\begin{figure}[ht]
\centering
\includegraphics[width=0.9\textwidth]{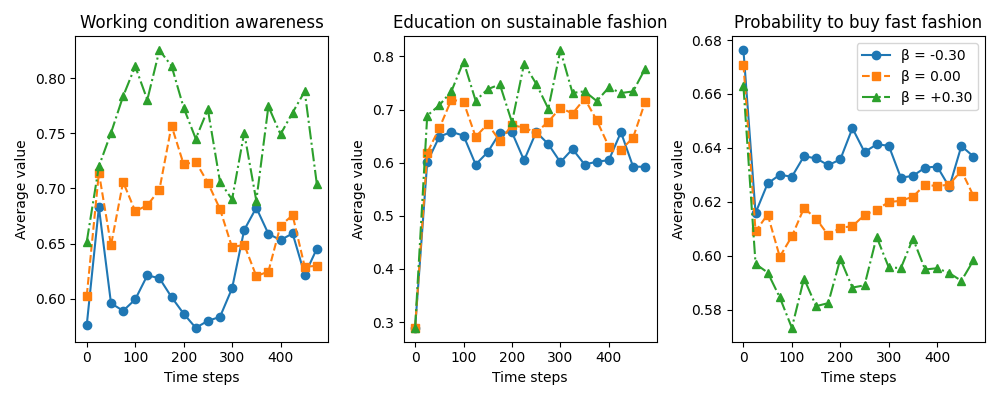}
\caption{Changes in average values for Model (C3)
\textit{Source: our simulations.}}
\label{fig:graphA9}
\end{figure}

\begin{figure}[ht]
\centering
\includegraphics[width=0.9\textwidth]{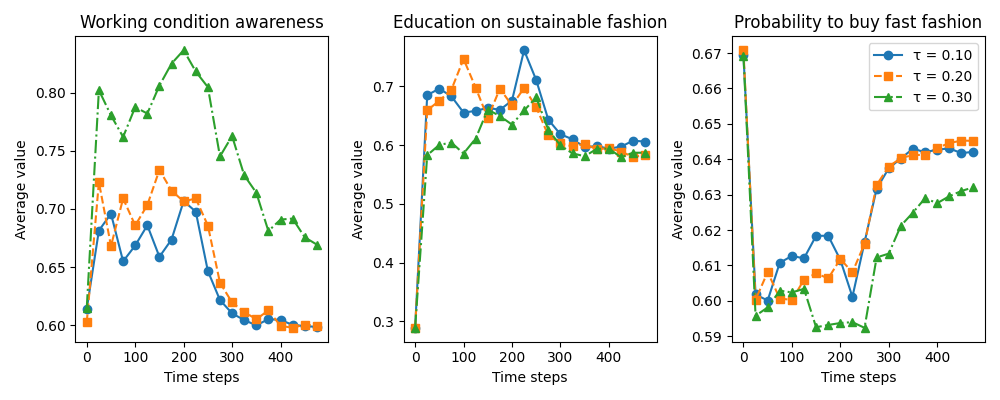}
\caption{Changes in average values for Model (C4)
\textit{Source: our simulations.}}
\label{fig:graphA10}
\end{figure}

\end{document}